\documentclass[aps,prb,10pt,twocolumn,english,superscriptaddress,citeautoscript,showkeys,preprintnumbers,amsmath,amssymb,floatfix,footinbib]{revtex4-2}
\usepackage[utf8]{inputenc}
\usepackage{hyperref}
\hypersetup{colorlinks=true,linkcolor=red,citecolor=blue}
\usepackage{amsmath}
\usepackage{amssymb}
\usepackage{graphicx}
\usepackage{dcolumn}
\usepackage{xcolor}
\usepackage{float}
\usepackage{soul}
\usepackage{adjustbox}
\usepackage{lipsum}
\usepackage{blindtext}
\usepackage{lipsum}
\usepackage{siunitx}

\usepackage{placeins}

\mathchardef\mhyphen="2D

\begin{document}

\title{Microscopic origin of hard-plane antiferromagnetism in the Kondo lattice \texorpdfstring{Ce$_2$Rh$_3$Ge$_5$}{Ce2Rh3Ge5}}

\author{Rajesh Tripathi}
\email{raj7tpi@gmail.com}
\affiliation{Department of Physics, Indian Institute of Technology (BHU) Varanasi, 221005 Uttar Pradesh, India}
\affiliation{ISIS Neutron and Muon Source, STFC, Rutherford Appleton Laboratory, Chilton, Oxon OX11 0QX, United Kingdom}
\author{Ewan Scott}
\email{ewan.scott.18@ucl.ac.uk}
\affiliation{
Department of
Mathematics, University College London, London WC1H 0AY, UK}
\author{D. T. Adroja}
\email{devashibhai.adroja@stfc.ac.uk}
\affiliation{ISIS Neutron and Muon Source, STFC, Rutherford Appleton Laboratory, Chilton, Oxon OX11 0QX, United Kingdom}
\affiliation{Highly Correlated Matter Research Group, Physics Department, University of Johannesburg, Auckland Park 2006, South Africa}

\author{D. Das}
\affiliation{Laboratory for Muon Spin Spectroscopy, Paul Scherrer Institute, CH-5232 Villigen PSI, Switzerland}
\author{C. Ritter}
\affiliation{Institut Laue-Langevin, 71 Avenue des Martyrs, CS 20156, 38042, Grenoble Cedex 9, France}

\author{Huanzhi Hu}
\affiliation{
Department of
Physics, University of Birmingham, Birmingham B15 2TT, UK}

\author{Michal P. Kwasigroch}
\email{m.kwasigroch@ucl.ac.uk}
\affiliation{Department of
Mathematics, University College London, London WC1H 0AY, UK} 
\affiliation {Trinity College, Cambridge CB2 1TQ, UK}
\author{Nicholas Corkill}
\affiliation{Department of Physics, Kent State University, Kent, OH 44242, USA}
\author{Gheorghe Lucian Pascut}
\affiliation{Functional Materials \& Sustainability Laboratory, Faculty of Forestry, Stefan Cel Mare University (USV), Suceava 720229, Romania}
\affiliation{Crystal Growth Laboratory, Faculty of Physics, West University of Timisoara, 4 Bd. Vasile Parvan,
300223 Timisoara, Romania}
\author{T. Masuda}
\affiliation{Institute for Solid State Physics, University of Tokyo, Kashiwa, Chiba 277-8581, Japan}
\affiliation{Institute of Materials Structure Science, High Energy Accelerator Research Organization, Ibaraki 305-0801, Japan}
\affiliation{Trans-scale Quantum Science Institute, The University of Tokyo, Tokyo 113-0033, Japan}

\author{S. Asai}
\affiliation{Institute for Solid State Physics, University of Tokyo, Kashiwa, Chiba 277-8581, Japan}
\author{T. Takabatake}
\affiliation{Department of Quantum Matter, Graduate School of Advanced Science and Engineering, Hiroshima University, Higashi-Hiroshima, 739-8530, Japan}
\author{T. Onimaru}
\affiliation{Department of Quantum Matter, Graduate School of Advanced Science and Engineering, Hiroshima University, Higashi-Hiroshima, 739-8530, Japan}

\author{T.~Shiroka}
\affiliation{PSI Center for Neutron and Muon Sciences CNM, CH-5232 Villigen PSI, Switzerland}
\affiliation{Laboratorium f\"{u}r Festk\"{o}rperphysik, ETH Z\"{u}rich, CH-8093 Z\"{u}rich, Switzerland}
\author{Francis Pratt}
\affiliation{ISIS Neutron and Muon Source, STFC, Rutherford Appleton Laboratory, Chilton, Oxon OX11 0QX, United Kingdom}
\author{A. M. Strydom}
\affiliation{Highly Correlated Matter Research Group, Physics Department, University of Johannesburg, Auckland Park 2006, South Africa}
\author{S. Langridge}
\affiliation{ISIS Neutron and Muon Source, STFC, Rutherford Appleton Laboratory, Chilton, Oxon OX11 0QX, United Kingdom}
\author {A. Sundaresan}
\affiliation{School of Advanced Materials and Chemistry and Physics of Materials Unit, Jawaharlal Nehru Centre for Advanced Scientific Research, Jakkur, Bangalore 560064, India}
\author {S. Patil}
\affiliation{Department of Physics, Indian Institute of Technology (BHU) Varanasi, 221005 Uttar Pradesh, India}

\date{\today}

\begin{abstract}

Hard-plane antiferromagnetic order — where ordered moments lie perpendicular to the single-ion crystal electric field easy axis — is rare in Ce-based Kondo lattices and is a subject of active interest. Here we show that Ce$_2$Rh$_3$Ge$_5$ realizes a hard-plane antiferromagnetic state in which partial delocalization of the local moment gives rise to an RKKY exchange that overturns the single-ion easy-axis preference.
Neutron diffraction reveals moments in the $ab$ plane, while inelastic neutron scattering and susceptibility establish a magnetic easy axis along $c$ in the paramagnetic regime, highlighting a clear inversion between single-ion and ordered-state anisotropies.
In this work, we establish a unified microscopic framework to consistently account for partial $4f$-moment delocalization, enhanced in-plane RKKY exchange, and the resulting hard-plane antiferromagnetic order. Ce$_2$Rh$_3$Ge$_5$ thus provides a benchmark system in which single-ion anisotropy, Kondo screening, and RKKY exchange compete on comparable energy scales, revealing a cooperative route to hard-axis ordering in strongly hybridized Kondo lattices.

\end{abstract}


\maketitle
\label{Intro}

Magnetic anisotropy in rare-earth intermetallics is normally dictated by the crystalline electric field (CEF), which lifts the degeneracy of the $4f$ multiplet and selects a single-ion easy axis or plane along which moments preferentially align. In Ce-based Kondo lattices this single-ion anisotropy can compete with the Ruderman--Kittel--Kasuya--Yosida (RKKY) exchange interaction, which promotes intersite magnetic order. On top of this competition, the Kondo effect delocalizes the local moments and renormalizes the magnetic response. The intricate balance between these energy scales underlies a broad range of heavy-fermion phenomena including unconventional superconductivity, non-Fermi-liquid behavior, quantum criticality, and strongly reduced ordered moments~\cite{RevModPhys.73.797,RevModPhys.79.1015,doi:10.1126/science.1191195,doi:10.1126/science.aaa9733,PhysRevLett.43.1892,mathur1998magnetically}.

While the CEF single ion anisotropy usually controls the direction of the ordered moment, an intriguing class of Ce- and Yb-based compounds exhibit \emph{hard-axis} or \emph{hard-plane} magnetic order, in which the ordered moments lie perpendicular to the single-ion easy direction inferred from the high-temperature susceptibility~\cite{PhysRevB.104.115169,PhysRevB.91.035102,PhysRevB.94.014418,Krellner_2011}. This anomalous anisotropy has been primarily observed in heavy-fermion ferromagnets such as CeAgSb$_2$~\cite{PhysRevB.104.115169}, YbNi$_4$P$_2$~\cite{Krellner_2011}, and Ce$_2$PdGe$_3$~\cite{PhysRevB.94.014418,PhysRevB.91.035102} and has motivated two distinct lines of explanation. One attributes the effect to \emph{anisotropic RKKY exchange}, shaped by crystal symmetry, Ce–Ce geometry, and conduction-band anisotropy, which can overcome the CEF preference at the level of the collective order parameter~\cite{Hanshang,PhysRevB.82.100405}. The other proposes that strong Kondo hybridization itself renormalizes the exchange anisotropy, favoring magnetic order along a CEF hard direction in the 2-fluid regime where local-moment and heavy-fermion quasiparticles coexist~\cite{Hanshang,Scott2025}. In particular, recent numerical calculations~\cite{wojcik2026} on a single magnetic impurity have shown that an inversion of the anisotropy can take place at intermediate temperatures where the local moment is only partially screened -- this is the analogue of the 2-fluid regime in a Kondo lattice.

To date, however, such hard-axis/plane ordering has been documented predominantly in \emph{ferromagnetic} Ce- or Yb-based Kondo systems, often accompanied by multiple transitions~\cite{PhysRevB.104.115169,Krellner_2011,PhysRevB.94.014418,PhysRevB.91.035102,Hanshang,PhysRevB.82.100405}, magnetic instabilities, or proximity to quantum criticality. Whether analogous behavior can emerge in \emph{antiferromagnetic} Ce- or Yb-based Kondo lattices, and if so by what microscopic mechanism, remains an open question of broad relevance for understanding how CEF physics, anisotropic exchange, and Kondo hybridization conspire to select ordered states in correlated $f$-electron materials.

The ternary Ce$_2T_3X_5$ ($T =$ transition metal, $X =$ Si or Ge), so-called 235 family, provides a fertile arena for addressing this question, hosting a rich variety of correlated ground states that arise from comparable CEF, RKKY, and Kondo energy scales~\cite{Brown2023,GODART1988677,PhysRevB.60.10383,PhysRevB.69.174423,Nakashima2005,WINIARSKI2013123,SEGRE1981372}. Small changes in $f$-$d$ hybridization or Ce–Ce geometry drive members toward antiferromagnetism, valence fluctuations, non-Fermi-liquid behavior, or even pressure-induced superconductivity, attesting to their proximity to magnetic instability and electronic reconstruction.

Within this family, Ce$_2$Rh$_3$Ge$_5$ stands out as an intriguing Kondo lattice antiferromagnet. Bulk probes indicate antiferromagnetic order at $T_{\mathrm{N}}\approx 5.5$\,K and a Kondo scale at least an order of magnitude higher~\cite{PhysRevB.60.10383,PhysRevB.64.144412}, suggesting significant Kondo renormalization and competition with intersite exchange. Pressure suppresses the magnetic order and induces quantum critical behavior beyond $P_{\mathrm{c}}\sim 0.45$\,GPa~\cite{PhysRevB.64.144412}, placing Ce$_2$Rh$_3$Ge$_5$ near a putative quantum critical point. Yet the microscopic magnetic structure, CEF scheme, and internal-field landscape have remained unresolved, precluding any discussion of whether its ordering anisotropy follows conventional CEF expectations or reflects the more exotic exchange- or Kondo-driven mechanisms proposed for hard-direction systems.

Here we combine neutron powder diffraction (NPD), inelastic neutron scattering (INS), and muon spin rotation/relaxation ($\mu$SR) with microscopic modeling of the CEF Hamiltonian and an anisotropic Coqblin--Schrieffer framework to determine the magnetic structure, CEF level scheme, and internal field environment of Ce$_2$Rh$_3$Ge$_5$. We supplement our low-energy theory with isotropic density functional theory plus embedded dynamical mean-field theory (DFT+eDMFT) calculations at moderate temperatures that reveal a Kondo temperature that is comparable with the CEF energy scale and partial delocalization of the local moments that coexist with the heavy-fermion quasiparticles. The above multimodal approach reveals that Ce$_2$Rh$_3$Ge$_5$ hosts a \emph{hard-plane} antiferromagnetic ground state with strongly reduced ordered moments, a single-ion CEF easy $c$ axis, and an RKKY exchange that is strongest in the $ab$ plane. 

Here, we show that anisotropic exchange and Kondo hybridization cooperate to reverse the single-ion anisotropy at the level of the ordered state, thereby extending the hard-axis/plane phenomenology into the antiferromagnetic sector of Kondo lattices.

\section*{Results}

\subsection*{Neutron diffraction study}

\begin{figure*}
		\includegraphics[width=\linewidth]{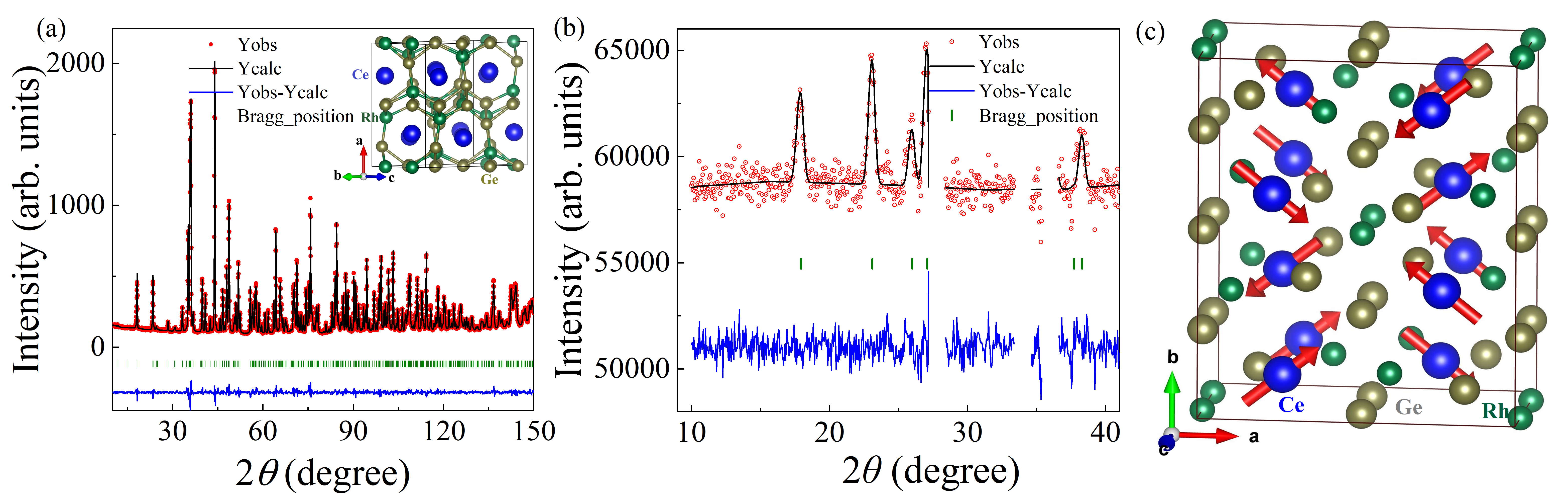}
		\caption{\textbf{Magnetic structure revealed by neutron diffraction:} (a) Rietveld-refined NPD pattern of Ce$_2$Rh$_3$Ge$_5$ measured at room temperature on the D2B diffractometer at ILL using a wavelength $\lambda = 1.594$\,\AA. Red circles represent the experimental data, the black line is the calculated pattern for the orthorhombic U$_2$Co$_3$Si$_5$-type structure, green vertical ticks mark the Bragg reflection positions, and the blue line at the bottom shows the difference between observed and calculated intensities. The inset illustrates the crystal structure, emphasizing the local coordination around the Ce site within the three-dimensional Rh--Ge framework. The blue, green, and gray spheres represent Ce, Rh, and Ge atoms, respectively. (b) Refinement of the temperature-difference diffraction pattern (1.6\,K-10\,K), where a constant offset of 60000 counts was added to shift all the data to positive values. The scale factor was fixed based on the refinement of the paramagnetic phase at 10\,K. The solid black line represents the calculated magnetic intensity assuming a propagation vector $\mathbf{k} = (0, 0, 0)$ with AFM coupling in the $ab$ plane, as depicted in panel (c). The blue line indicates the difference between observed and calculated intensities, and green vertical ticks denote the positions of magnetic Bragg reflections. (c) Crystal and magnetic structure of Ce$_2$Rh$_3$Ge$_5$, showing AFM alignment of Ce moments along both $a$ and $b$ axes. No ordered magnetic moment component along the $c$ axis is detected within experimental uncertainty.}
		\label{ND}	
\end{figure*}

The high-resolution powder data for Ce$_2$Rh$_3$Ge$_5$ from D2B confirm the U$_2$Co$_3$Si$_5$-type orthorhombic structure; Table~\ref{table1} gives the refined values of the lattice parameters and the atom positions. As shown in Fig.~\ref{ND}(a), the refinement confirms the single-phase nature of the sample and the lattice parameters are in good agreement with the previous report~\cite{PhysRevB.60.10383}. The high-intensity NPD data taken on D20 reveal weak additional magnetic intensity superimposed on nuclear reflections as well as very small extra reflections in the 1.6~K and 4~K data upon subtracting the data taken in the paramagnetic state at 10~K. This additional coherent intensity was attributed to magnetic scattering and using the program K-Search, part of the Fullprof suite of programs~\cite{RODRIGUEZCARVAJAL199355}, the relevant reflections were indexed with a magnetic propagation vector $\mathbf{k} = 0$. Magnetic symmetry analysis using the program Basireps~\cite{RodriguezFullProf,ritter2011} was done for Ce on the Wyckoff site $8j$ of space group $Ibam$. There exist eight allowed irreducible representations (IR). Four IRs have only one basis vector (BV) in the direction of orthorhombic $c$-axis, while the remaining four each possess two BVs describing the coupling in directions of the $a$ and $b$ axes. Of the eight IRs, only one reproduces the magnetic Bragg reflections satisfactorily, see Fig.~\ref{ND}(b).

\begin{table}
	\centering
	\caption {Room-temperature atomic coordinates of Ce$_2$\-Rh$_3$\-Ge$_5$ refined using the $Ibam$ space group: $a = 10.926(1)$\,\AA, $b=12.1026(1)$\,\AA, and $c=5.9937(1)$\,\AA. The sites occupancies were fixed to unity.}
	\vskip 2mm
    \renewcommand{\arraystretch}{1.4}
	\addtolength{\tabcolsep}{+8.5pt}
	\begin{tabular}{c c c c c }
		\hline
		\hline
	    Atom & Site   & $x$        &  $y$         & $z$      \\ \hline
		Ce   & $8j$   & 0.2675(2)  &  0.1334(2)   & 0        \\
		Rh1  & $4a$	  & 0          &   0          &  1/4     \\
		Rh2  & $8j$   & 0.1053(1)  &  0.3596(1)   & 0        \\
		Ge1  & $4b$   & 1/2        &  0           & 1/4      \\
		Ge2  & $8g$	  & 0          &  0.2224(1)   & 1/4      \\
		Ge3  & $8j$	   & 0.3392(1) &  0.4005(1)   & 0        \\
		\hline
		\hline
	\end{tabular}
	\label{table1}
\end{table}

To determine the magnetic structure, we used temperature-difference data (1.6–10~K and 4–10~K), since the magnetic signal is very weak ($<$0.1\% of total scattering). The difference patterns, which contain only magnetic intensity, were refined using a fixed scale factor obtained from the nuclear refinement of the 10~K paramagnetic data. The magnetic structure obtained from the refinement corresponds to an AFM coupling in the $a$ and $b$ direction with no component in the $c$-direction (Fig.~\ref{ND}(c)), magnetic space group $Ibam'$. At 1.6~K the refinement generates the magnetic components in $a$- and $b$-direction: $\mu_x$ = 0.24(1)~$\mu_{\text{B}}$, $\mu_y$ = 0.19(1)~$\mu_{\text{B}}$ and a total magnetic moment $\mu_{\text{Ce}}$ = 0.31(1)~$\mu_{\text{B}}$, which is at $38^\circ$ from the $a$ axis. At 4~K the magnetic components in $a$- and $b$-direction are reduced to: $\mu_x$ = 0.17(1)~$\mu_{\text{B}}$, $\mu_y$ = 0.15(1)~$\mu_{\text{B}}$ and the total magnetic moment to $\mu_{\text{Ce}}$ = 0.23(1)~$\mu_{\text{B}}$. The strong reduction of the ordered state Ce magnetic moment compared to the expected theoretical value of $g_{\text{J}}J$ = 2.14~$\mu_{\text{B}}$ for a free trivalent Ce ion indicates the presence of strong Kondo screening and CEF effects, which substantially reduce the static ordered moment. 

To place the magnetic structure of Ce$_2$Rh$_3$Ge$_5$ in context, it is instructive to compare it with other Ce-235 compounds crystallizing in the U$_2$Co$_3$Si$_5$-type structure. For example, Ce$_2$Ru$_3$Ge$_5$ exhibits a similar $\mathbf{k} = 0$ AFM order with $T_\mathrm{N} = 4.1$~K, but with a larger ordered moment ($\sim$0.5–0.7~$\mu_\mathrm{B}$), indicative of weaker Kondo compensation~\cite{doi:10.1021/acs.inorgchem.5c01691}. In contrast, Ce$_2$Ni$_3$Ge$_5$ undergoes two successive magnetic transitions ($T_\mathrm{N} = 5$~K, $T^\ast = 4.3$~K), forming a canted AFM structure with propagation vector $\mathbf{q} = [0, 1, 0]$ and strongly anisotropic moments ($\mu_a = 0.45$~$\mu_\mathrm{B}$, $\mu_b = 0.15$~$\mu_\mathrm{B}$), tilted $\sim 20^\circ$ from the \textit{a} axis~\cite{HONDA2008504,doi:10.1143/JPSJ.74.2843}.

Among these U$_2$Co$_3$Si$_5$-type compounds, Ce$_2$Rh$_3$Ge$_5$ is notable for its minimal ordered moment ($\mu_\mathrm{Ce} \approx 0.31~\mu_\mathrm{B}$ at 1.6~K) and robust $\mathbf{k} = 0$ AFM order, reflecting the combined influence of strong 4$f$–conduction electron hybridization and CEF effects. While its magnetic anisotropy closely resembles that of Ce$_2$Ru$_3$Ge$_5$ and Ce$_2$Ni$_3$Ge$_5$, the substantially reduced moment highlights the significant role of Kondo screening in this system.

\subsection*{Inelastic neutron scattering study}

\begin{figure*}
		\includegraphics[width=14cm, keepaspectratio]{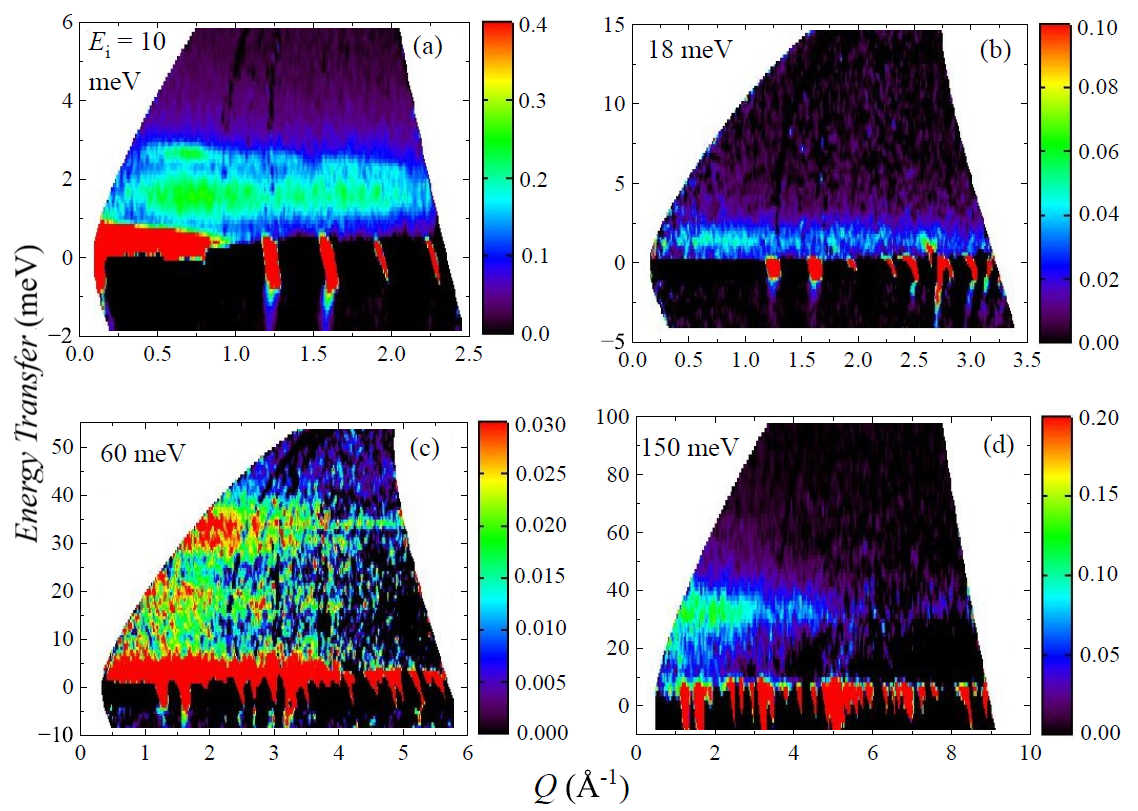}
		\caption{\textbf{Crystal field excitations revealed by inelastic neutron scattering:} Color-coded INS intensity maps of Ce$_2$Rh$_3$Ge$_5$ after phonon subtraction using La$_2$Rh$_3$Ge$_5$ at 4~K, plotted as a function of energy and momentum transfer, measured with incident neutron energies of (a) $E_i$ = 10~meV, (b) 18~meV, (c) 60~meV, and (d) 150~meV. Intensities are shown in arbitrary units.}
		
		\label{INS1}
\end{figure*}

\begin{figure*}
		\includegraphics[width=16.5 cm, keepaspectratio]{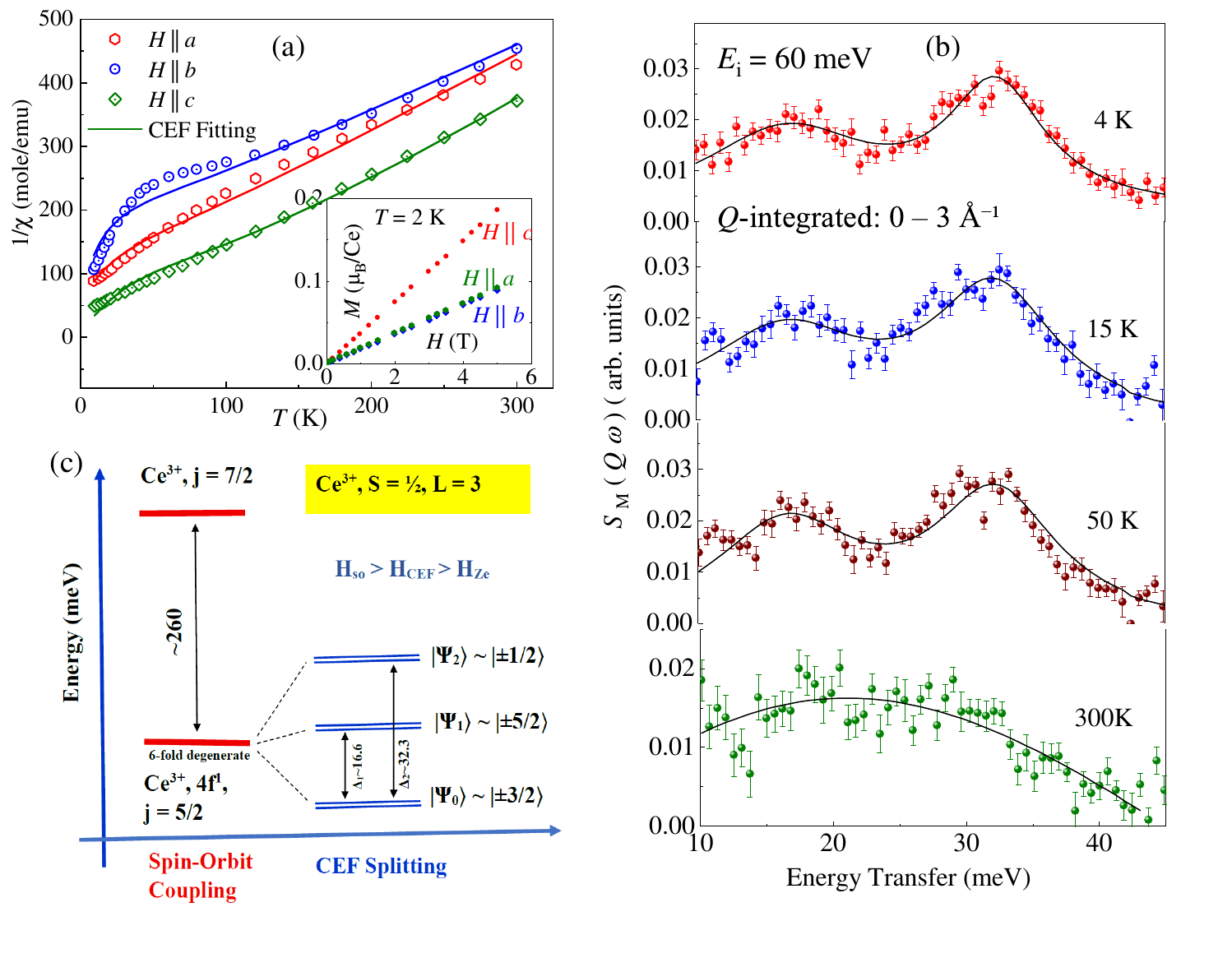}
		\caption{\textbf{Crystal electric field Hamiltonian - modeling and fits to experimental data:} (a) Inverse magnetic susceptibility of single-crystal Ce$_2$Rh$_3$Ge$_5$ measured at $H = 0.1$~T along the $a$, $b$, and $c$ axes~\cite{PhysRevB.64.144412}. Solid curves represent fits to the CEF model. Inset: Magnetization isotherms measured along the three principal crystallographic directions at $T = 2$~K. (b) Magnetic INS response after phonon subtraction using La$_2$Rh$_3$Ge$_5$, measured with $E_i = 60$~meV and $Q$-integrated over $0$-$3$~$\text{\AA}^{-1}$ at selected temperatures in the paramagnetic state. Solid curves are fits to the CEF model. (c) Schematic CEF level scheme of Ce$^{3+}$ ($J$=5/2) showing the splitting into three Kramers doublets.}
		
		\label{INS2}
\end{figure*}

High-energy INS measurements on Ce$_2$Rh$_3$Ge$_5$ were performed to investigate the CEF excitations of the $4f$ electrons and the wave functions of the CEF levels. To isolate the magnetic contribution in Ce$_2$Rh$_3$Ge$_5$, the phonon background was subtracted using INS data from the corresponding nonmagnetic isostructural compound La$_2$Rh$_3$Ge$_5$, scaled by a factor $\alpha$ = 0.87, based on the ratio of the total scattering cross sections. However, residual phonon intensity remained at high momentum transfer ($Q$), indicating an underestimation of the phonon background. Refining the scaling factor to $\alpha = 1.1$ yields improved subtraction results, as shown in Figs.~\ref{INS1}(a)–(d).

In the difference spectra at $E_i$ = 10~meV and $T$ = 4~K (Fig.~\ref{INS1}(a)), two low-energy excitations at approximately 1.7 and 2.5~meV are observed. While the 1.7~meV excitation is attributed to magnetic spin-wave modes below the N\'eel temperature, the second feature at 2.5~meV shows no temperature dependence and is absent in the $E_i$ = 18~meV data (Fig.~\ref{INS1}(b)), suggesting it may arise from spurious multiple scattering, possibly from the cold cryostat (CCR) walls. This 2.5~meV signal was also observed at 10 and 20~K (i.e. above $T_{\text{N}}$) (see Supplementary Fig.~1(a)), further confirming the spurious multiple scattering.

At higher energies, two distinct magnetic excitations around 16 and 32~meV are clearly visible in the $E_i$ = 60~meV data (Fig.~\ref{INS1}(c)), corresponding to transitions between the CEF-split states of Ce$^{3+}$ ($J = 5/2$). These excitations are consistent with the splitting of the sixfold degenerate ground state into three Kramers doublets due to the orthorhombic crystal field (Fig.~\ref{INS2}(c)). Additional measurements with $E_i$ = 150~meV (Fig.~\ref{INS1}(d)) show no magnetic scattering above 40~meV, confirming that the total CEF splitting lies below this energy.

To quantitatively analyze the magnetic excitations, we modeled the CEF using the Stevens operator formalism, assuming orthorhombic point symmetry ($mm2$, or C$_{2v}$) at the Ce site. The corresponding Hamiltonian is

\begin{equation}
H_\mathrm{CEF} = B_{2}^{0}O_{2}^{0} +B_{2}^{2}O_{2}^{2}+B_{4}^{0}O_{4}^{0} + B_{4}^{2}O_{4}^{2}+ B_{4}^{4}O_{4}^{4},
\label{H-CEF}
\end{equation}

where $B_n^m$ are CEF parameters and $O_n^m$ are Stevens operators~\cite{Stevens1952}. The $c$-axis (parallel to $z$ axis) was taken as the quantization axis. The CEF parameters were obtained by simultaneously fitting the magnetic INS data and single-crystal magnetic susceptibility data~\cite{PhysRevB.64.144412} using the Mantid software suite~\cite{ARNOLD2014156}, with a Lorentzian line shape applied to the inelastic excitations.

The solid curves in Fig.~\ref{INS2}(a) and Fig.~\ref{INS2}(b) represent the fits to the susceptibility and INS data, respectively. The best-fit CEF parameters (in meV) are: $B_2^0$ = --0.6500(2), $B_2^2$ = --0.5390(7), $B_4^0$ = 0.0870(8), $B_4^2$ = --0.02376(5), and $B_4^4$ = --0.1900(6). These values yield excited CEF doublets at $\Delta_1$ = 16.6~meV and $\Delta_2$ = 32.3~meV. The fitted molecular field constants $\lambda$ (in mol/emu) and temperature-independent susceptibilities $\chi_0$ (in emu/mol) for fields along the $a$, $b$, and $c$ directions are: $\lambda$ = --54(1), --47(2), and --6.6(7); $\chi_0$ = --0.9(6) $\times$ 10$^{-5}$, 1.4(7) $\times$ 10$^{-4}$, and --1.8(1) $\times$ 10$^{-4}$, respectively. 

The CEF analysis reflects the single-ion anisotropy of Ce$^{3+}$ in the high-temperature paramagnetic regime. Using the refined CEF parameters, we calculated the eigenfunctions of the three Kramers doublets, shown below. 

\begin{equation}
 \begin{split}
 \Psi_0 & = -0.077\, | \pm \frac{1}{2}\rangle +0.940\,|\mp \frac{3}{2} \rangle -0.331\,|\pm \frac{5}{2} \rangle   \\
 \Psi_1 & = 0.106 \,|\pm \frac{1}{2}\rangle -0.337 \,|\mp \frac{3}{2} \rangle +0.933 \, |\pm \frac{5}{2} \rangle \\ 
 \Psi_2 & = 0.991 \,|\pm \frac{1}{2}\rangle -0.037 \,|\mp \frac{3}{2} \rangle -0.126 \, |\pm \frac{5}{2} \rangle, \nonumber
 \end{split}
 \label{E-Fun}
\end{equation}

Given that both excited CEF levels lie at energies ($\Delta_{1} \approx 16.6~\mathrm{meV}$ $\approx 190~\mathrm{K}$, $\Delta_{2} \approx 32.3~\mathrm{meV} \approx 375~\mathrm{K}$) far above the magnetic ordering temperature, the low-temperature susceptibility and field-induced magnetization are governed entirely by the ground-state doublet. The ground-state Kramers doublet is dominated by the $\lvert \mp 3/2\rangle$ component, which carries a large $J_{z}$ projection and therefore produces a stronger magnetic response for $H \parallel c$. This establishes the $c$ axis as the single-ion CEF easy axis in the paramagnetic regime, fully consistent with both the experimental and the calculated susceptibility above $T_{\mathrm{N}}$, where $\chi_{c} > \chi_{a}, \chi_{b}$ (Fig.~\ref{INS2}(a) and Supplementary Fig.~1(c), respectively). 

The most intriguing aspect is the inversion between the single-ion anisotropy and the ordered moment direction. Remarkably, below $T_{\mathrm{N}}$, NPD reveals that the ordered moments lie within the $ab$ plane, and this is mirrored in the susceptibility: $\chi(T)$ for $H \parallel a$ and $H \parallel b$ decreases sharply below $T_{\mathrm{N}}$, as expected for longitudinal components parallel to the ordered moments, whereas $\chi(T)$ for $H \parallel c$ remains nearly temperature independent and thus behaves as a transverse susceptibility~\cite{PhysRevB.107.104412}. At the same time, the isothermal magnetization at 2~K still reflects the underlying single-ion CEF anisotropy. Even in the ordered phase, $M(H \parallel c)$ remains approximately twice as large as $M(H \parallel a)$ and $M(H \parallel b)$ up to 5~T, indicating that it is still easiest to polarize the Ce moments along the $c$ axis when a field is applied (inset of Fig.~\ref{INS2}(a) and Supplementary Fig.~1(d)). 
In other words, the CEF ground state continues to favor $c$-axis polarization for field-induced moments, while the anisotropic RKKY exchange stabilizes a static AFM order parameter confined to the $ab$ plane. This dichotomy between the axes selected by the single-ion response and by the collective order parameter is the hallmark of the hard-plane AFM state realized in Ce$_2$Rh$_3$Ge$_5$. This hard-plane AFM state arises because anisotropic RKKY exchange interactions between Ce sites overpower the single-ion CEF preference at low temperatures and favor a particular pairwise alignment that minimizes exchange energy in the $ab$-plane.

Hard-plane ordering has been reported in a limited number of Ce-, Pr-, and Yb-based intermetallic systems. For example, CeRh$_2$Al$_{10}$ orders along the orthorhombic $c$-axis, identified as the CEF hard-axis, with a strongly reduced moment of 0.34(2)\,$\mu_{\mathrm{B}}$~\cite{PhysRevB.82.100405}. Similarly, hard-plane magnetic ordering has been observed in Ce$_2$PdGe$_3$~\cite{PhysRevB.91.035102, PhysRevB.94.014418}. Recent studies on Ce-based Kondo lattice show that the magnetic ordering direction is determined not only by the CEF easy axis but also by the anisotropy of exchange interactions~\cite{Hanshang,PhysRevB.105.224418}. Notably, while CeCuSi exhibits easy-plane ferromagnetic order consistent with its CEF ground state, CeAgSb$_2$ orders along a CEF hard direction, due to competing anisotropic exchange~\cite{Hanshang, PhysRevB.104.115169}. This comparison illustrates a general mechanism whereby exchange anisotropy can override single-ion CEF preferences, providing a framework to understand the hard-plane AFM order in Ce$_2$Rh$_3$Ge$_5$.

\subsection*{\label{muSR} Muon Spin Relaxation: Zero-Field \texorpdfstring{$\mu$SR}{muSR} Measurements}

\begin{figure*}
  \centering
		\includegraphics[width=14.5 cm, keepaspectratio]{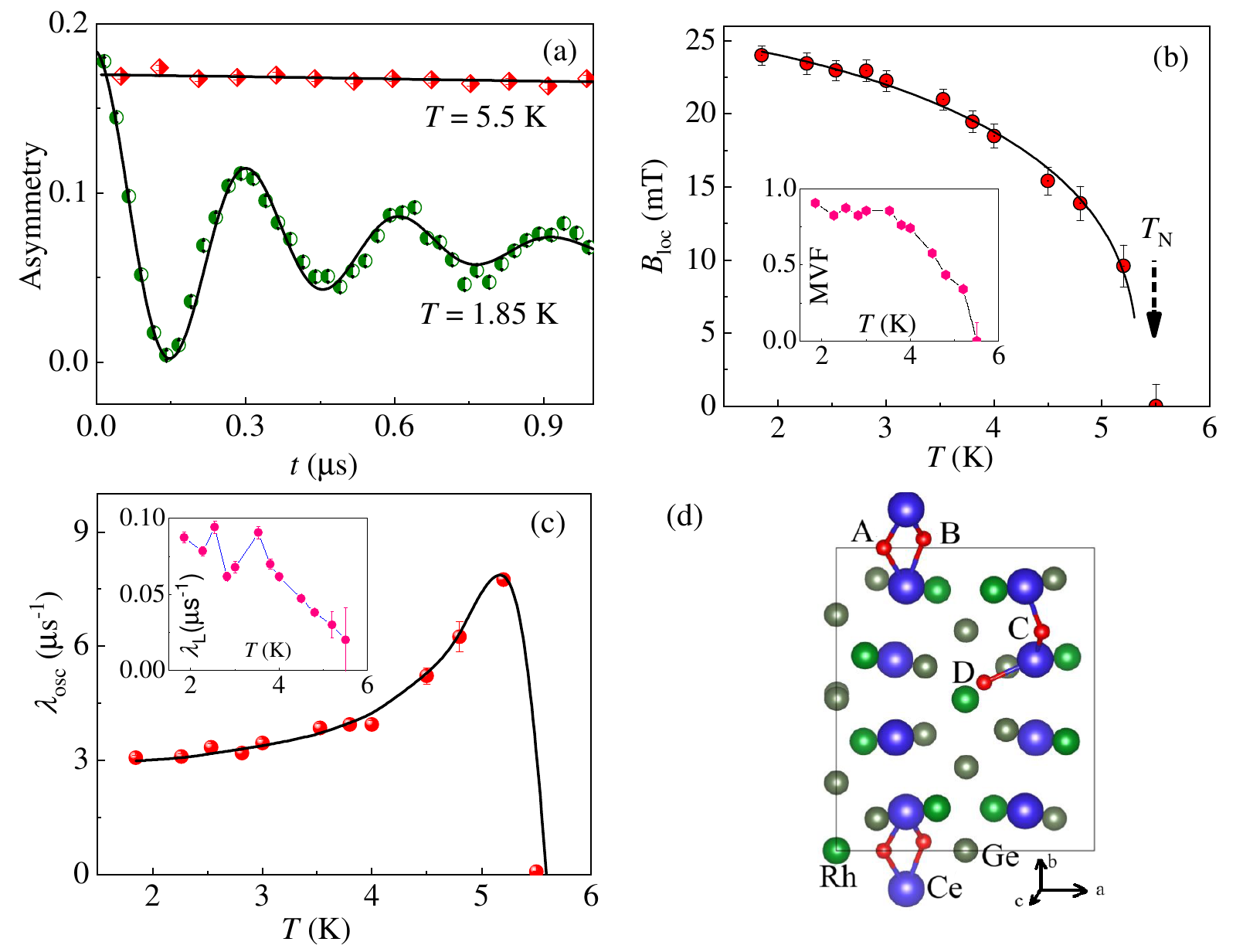}
		\caption{\textbf{\label{musr}Zero-field $\mu$SR results for Ce$_2$Rh$_3$Ge$_5$:} (a) ZF-$\mu$SR spectra recorded at 1.85 and 5.5\,K. Symbols represent experimental data, while the solid red lines show fits using Eq.~(\ref{eq:musr1}). (b) Temperature dependence of the internal local field $B_{\text{loc}}$; the solid line shows a phenomenological fit using Eq.~(\ref{eq:musr2}). The inset shows the magnetic volume fraction (MVF) derived from the oscillating asymmetry. (c) Transverse relaxation rate $\lambda_{\text{osc}}$, associated with the oscillatory component in Eq.~(\ref{eq:musr1}), vs temperature. Inset: Longitudinal relaxation rate $\lambda_{\text{L}}$ (associated with the non-oscillatory component in Eq.~(\ref{eq:musr1})) vs temperature. (d) Visualization of muon stopping sites within the crystallographic unit cell. The four potential muon stopping sites in this structure, marked A, B, C and D, are adopted from the isostructural compound Pr$_2$Pd$_3$Ge$_5$ Ref.~\cite{PhysRevB.107.104412}.}
\end{figure*}

To gain microscopic insight into the magnetic ordering in Ce$_2$Rh$_3$Ge$_5$, we performed ZF-$\mu$SR experiments over the temperature range 1.85--5.5\,K. Representative spectra at 1.85 and 5.5\,K are shown in Fig.~\ref{musr}(a). At 1.85\,K, the clear oscillations in the time-dependent asymmetry signal indicate the development of long-range magnetic order. In contrast, the 5.5~K spectrum shows a purely relaxing behavior without oscillations, indicating a paramagnetic state. The ZF-$\mu$SR data are well described by the following function:

\begin{equation}
A(t) = A_0 \left[ A_{\text{osc}} \cos(2\pi \nu t + \phi) e^{-\lambda_{\text{osc}} t} + (1 - A_{\text{osc}}) e^{-\lambda_L t} \right],
\label{eq:musr1}
\end{equation}

Here, $A_0$ represents the total initial asymmetry, while $A_{\text{osc}}$ denotes the fraction of $A_0$ associated with the oscillatory (transverse) component of the muon-spin polarization. The oscillatory signal reflects the coherent precession of muon spins in regions with static internal magnetic fields, with the corresponding precession frequency given by $\nu = \gamma_\mu B_{\text{loc}}$. Here, $\gamma_\mu = 135.53$\,MHz/T is the muon gyromagnetic ratio, while $B_{\text{loc}}$ represents the internal magnetic field at the muon site. The damping of the oscillatory component ($\lambda_{\text{osc}}$) reflects the distribution of static internal fields (dipolar broadening) in the ordered state. The remaining, non-oscillatory part corresponds to the longitudinal component, which arises from muon spins that are aligned with the internal magnetic field. This component relaxes exponentially with a rate $\lambda_L$ and probes low-frequency dynamic fluctuations. In the paramagnetic state, no spontaneous internal field is present on the $\mu$SR timescale, and hence the precessing component $A_{\mathrm{osc}}$ vanishes. The spectra are thus described solely by the longitudinal exponential relaxation term.

The initial asymmetry was fixed to $A_0 = 0.1837$, a value reliably determined from the high-temperature paramagnetic spectra, where there is no oscillatory component. This ensures a stable fitting procedure and prevents unnecessary parameter correlations. In a static polycrystalline AFM with full magnetic volume fraction, two-thirds of the muons experience transverse internal fields and contribute to the oscillatory asymmetry, while the remaining one-third form the longitudinal tail. We therefore quantify the magnetic volume fraction (MVF) as MVF($T$) = $A_{\text{osc}}$($T$)/(2/3). As shown in the inset of Fig.~\ref{musr}(b), the MVF increases sharply below $T_{\text{N}}$ and approaches $\sim$ 0.9 at the lowest temperatures, close to the ideal limit of unity expected for full magnetic volume. This behavior indicates that essentially the entire sample participates in the magnetic order and confirms the bulk and homogeneous nature of the magnetic phase~\cite{le2011muon}.

The temperature evolution of $B_{\text{loc}}$ is shown in Fig.~\ref{musr}(b). Below $T_{\text{N}}$, the field increases continuously and saturate at about 24\,mT at 1.85\,K, corresponding to a muon precession frequency of 3.1\,MHz. The observation of a dominant single precession frequency suggests that the $\mu$SR signal is governed by one principal muon stopping site. The temperature dependence of $B_{\text{loc}}$ below $T_{\text{N}}$ is further analyzed by fitting it to the phenomenological expression:

\begin{equation}
B_{\text{loc}}(T) = B_{\text{loc}}(0) \left[1 - \left( \frac{T}{T_{\text{N}}^{\mu}} \right)^{\delta} \right]^{\beta},
\label{eq:musr2}
\end{equation}

The fit is shown by the solid black curve in Fig.~\ref{musr}(b). This empirical model accounts for the suppression of the magnetic order parameter at low temperatures, commonly attributed to low-energy magnon excitations. In the fitting procedure, the critical exponent $\beta$ was fixed to the three-dimensional Heisenberg value ($\beta = 0.33$), consistent with the expected universality class for AFM ordering in Ce-based Kondo lattices and in line with previous $\mu$SR studies~\cite{PhysRevB.107.104412,PhysRevB.104.144405}. This constraint also stabilizes the fit, as $\beta$ and $T_{\text{N}}^{\mu}$ are strongly correlated parameters in $B_{\text{loc}}$($T$) fits. We obtain $B_{\text{loc}}(0) = 25.8(6)$~mT, consistent with the expected dipolar field scale for a Ce intermetallic, and $T_{\mathrm{N}}^{\mu} = 5.34(4)$\,K, marking the onset of quasistatic order on the $\mu$SR timescale. The exponent $\delta = 1.6(2) > 1$ reflects the additional low-temperature suppression of the ordered moment due to low-energy spin-wave (magnon) excitations which leads to a corresponding suppression of $B_{\mathrm{loc}}(T)$, a rationale commonly employed in this empirical description~\cite{DR1997,blundell2001}. 

Figure~\ref{musr}(c) shows the transverse relaxation rate $\lambda_{\text{osc}}$, which exhibits a pronounced maximum close to $T_{\text{N}}$, reflecting the combined effects of critical slowing down and the growth of static internal field inhomogeneities associated with the nucleation of long-range order. Such broadened relaxation peaks are typical in Kondo-lattice AFMs, where slow $4f$ spin dynamics and hybridization-driven fluctuations smear the transition relative to conventional local-moment systems~\cite{PhysRevB.85.134405,PhysRevB.99.224424,PhysRevB.104.144405}. In contrast, the longitudinal relaxation rate $\lambda_{\text{L}}$($T$) does not peak at $T_{\text{N}}$ but instead increases gradually upon cooling below the ordered state (inset of Fig.~\ref{musr}(c)), indicating the presence of low-frequency residual spin dynamics. This behavior is consistent with a partially screened Kondo lattice in which static AFM order coexists with slow fluctuations of the $4f$ moments~\cite{MACLAUGHLIN2003387}.

The ZF-$\mu$SR therefore confirms a bulk Kondo-screened AFM state with full magnetic volume fraction and slow residual spin dynamics, fully consistent with the reduced ordered moment from NPD. These findings are also in excellent agreement with the bulk signatures of magnetic ordering revealed by magnetic susceptibility and heat capacity~\cite{PhysRevB.60.10383, PhysRevB.64.144412}.

To further validate the internal field scale observed by $\mu$SR, we calculated the dipolar field values at various muon stopping sites using the magnetic structure model obtained from our NPD measurements. Density functional theory (DFT) calculations on the isostructural compound Pr$_2$Pd$_3$Ge$_5$ reveal four energetically favorable interstitial muon stopping sites (denoted A, B, C, and D)~\cite{PhysRevB.107.104412}. Given the close structural correspondence between the two compounds, these 
sites represent natural stopping positions also for Ce$_2$Rh$_3$Ge$_5$. As shown in Fig.~\ref{musr}(d), these correspond to crystallographic positions A = (0.184, 0.5, 0.25), B = (0.338, 0.523, 0.197), C = (0.796, 0.223, 0.306), and D = (0.573, 0.055, 0).

Using the ordered moment extracted from NPD, we evaluated the dipolar fields at the four sites and obtained 70~mT (A), 17~mT (B), 23~mT (C), and 40~mT (D). Since the internal field observed in the ZF-$\mu$SR experiments is approximately 25~mT, this matches closely the calculated field at site C. Notably, site C was previously found to be about 0.5~eV lower in energy than the other candidate sites in Pr$_2$Pd$_3$Ge$_5$, making it the most likely muon localization site also in Ce$_2$Rh$_3$Ge$_5$. This excellent agreement supports the interpretation that the $\mu$SR signal originates predominantly from muons stopping at site C, and that the commensurate in-plane magnetic structure derived from ND adequately describes the magnetic environment sensed by the muons.

\subsection*{Finite temperature calculations with Embedded Dynamical Mean Field Theory}

\begin{figure*}
  \centering
    \includegraphics[width=14.5 cm, keepaspectratio]{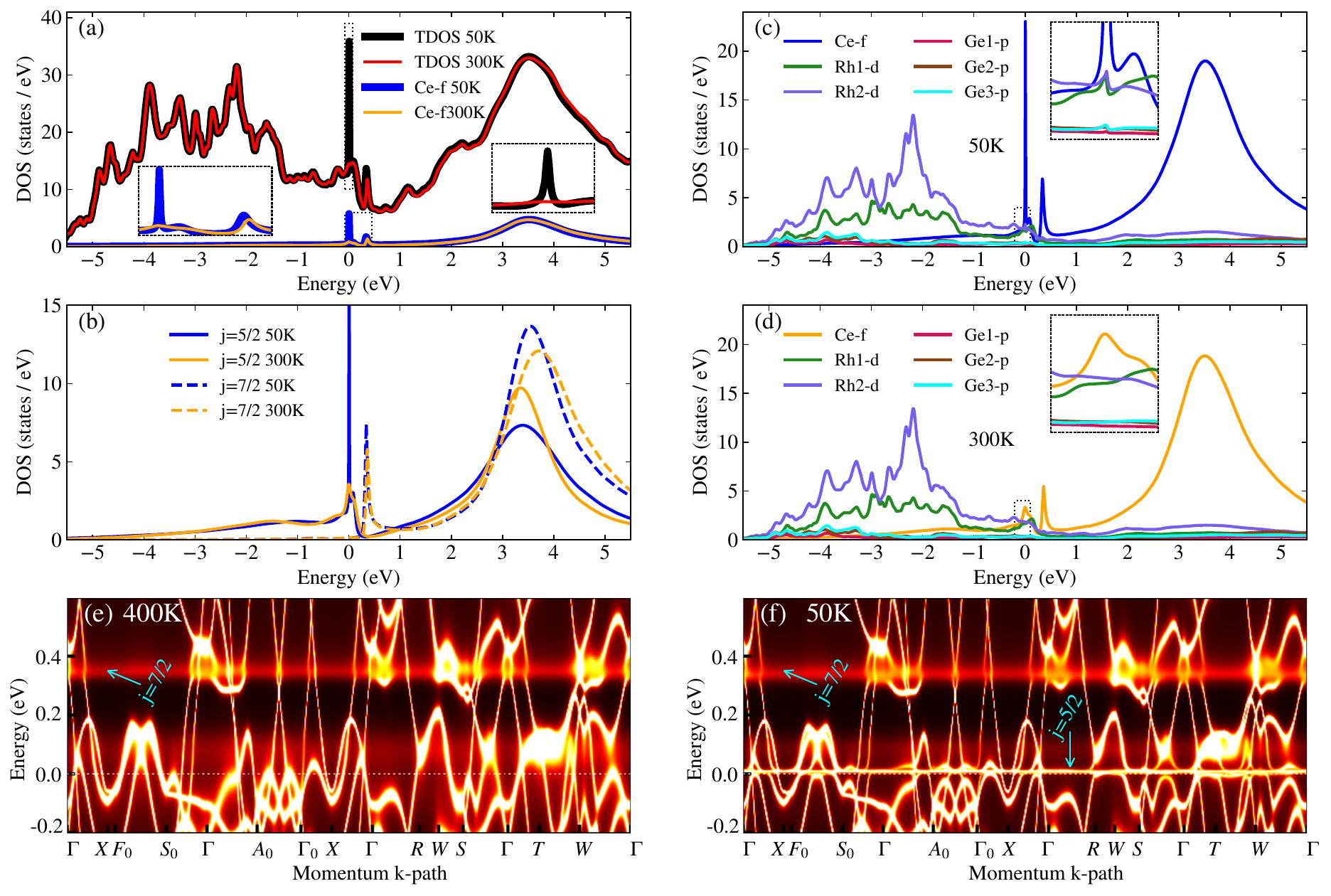}
		\caption{\textbf{Electronic properties of Ce$_2$Rh$_3$Ge$_5$ across different temperature regimes:} Panels (a)–(d) display the total, atomic, and orbital-projected density of states (DOS) per primitive unit cell. Panels (e) and (f) show the spectral functions in the vicinity of the Fermi level ($E_F$), with insets providing magnified views of the regions indicated by the dotted rectangles. In all panels, zero energy corresponds to the $E_F$, which is also marked by a horizontal dotted line in (e) and (f). The arrows in these panels highlight the Kondo resonances arising from the $j = 5/2$ and $j = 7/2$ electronic states.}
        \label{fig1T:eDMFT_DOS_PUC}
\end{figure*}

\begin{figure*}
  \centering
		\includegraphics[width=14.5 cm, keepaspectratio]{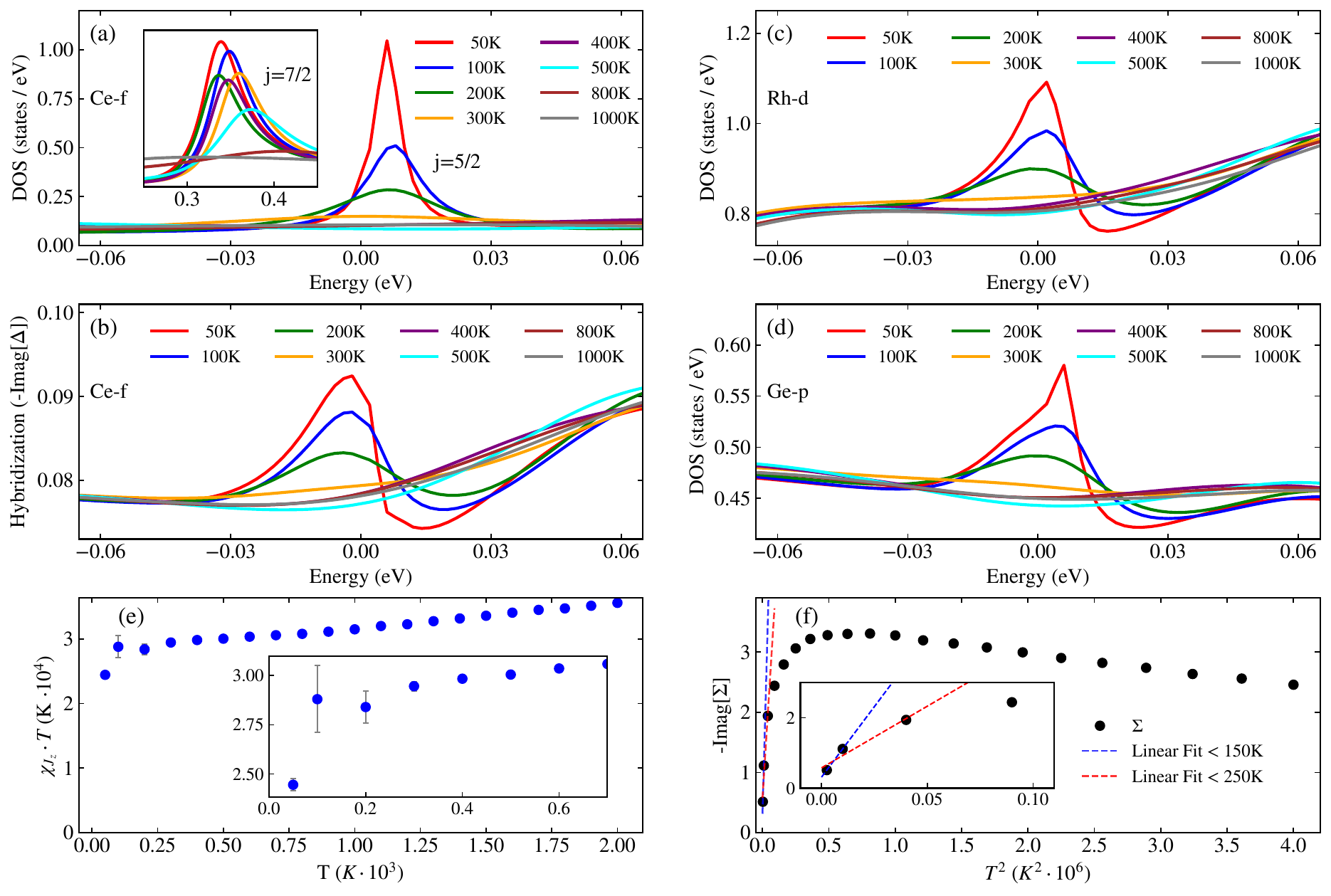}
		\caption{\textbf{Temperature-dependent electronic properties for Ce$_2$Rh$_3$Ge$_5$:} Panel (a) displays the Kondo resonance near the Fermi level ($E_F$), originating from the $j=5/2$ states, while the inset illustrates the behavior for the $j=7/2$ states. Panel (b) shows the corresponding hybridization function for the $j=5/2$ state. Panels (c) and (d) present the projected density of states (PDOS) for representative Rh and Ge ions within the unit cell; a qualitatively similar behavior is observed across all $s, p,$ and $d$ states for the remaining Rh and Ge ions. Panel (e) depicts the local magnetic susceptibility of the Ce $4f$ electrons, with the inset providing a detailed view of the low-temperature regime. Finally, panel (f) shows the imaginary part of the scattering rate for the $j=5/2$ state, where the inset zooms in on the low-temperature data. The dashed colored lines in the inset represent linear fits to these low-temperature results.}
        \label{fig2T:eDMFT_TDOS_PUC}
\end{figure*}

To investigate the isotropic electronic properties of Ce$_2$Rh$_3$Ge$_5$, we performed temperature-dependent electronic structure calculations using Density Functional Theory combined with embedded Dynamical Mean Field Theory (DFT+eDMFT).
As shown in Fig.~\ref{fig1T:eDMFT_DOS_PUC}(a), the total Density of States (DOS) remains largely consistent across a wide energy window at both 50~K and 300~K. However, a significant divergence occurs within a narrow region of $\pm 0.04$ eV around the Fermi level ($E_F$), where a sharp peak develops at 50~K. This peak originates from the Ce $f$-states—specifically the $j=5/2$ state, which is partially occupied by one electron, while the $j=7/2$ states remain empty (Fig.~\ref{fig1T:eDMFT_DOS_PUC}(b)). Furthermore, a detailed inspection of the $p$ and $d$ states of the Rh and Ge ions reveals that these states also develop peak-like features near $E_F$ as temperature decreases (Fig.~\ref{fig1T:eDMFT_DOS_PUC}(c)–(d)). These features are corroborated by the spectral functions shown in Fig.~\ref{fig1T:eDMFT_DOS_PUC}(e–f).
These low-temperature electronic signatures suggest the formation of a Kondo lattice in the Ce$_2$Rh$_3$Ge$_5$ compound. To further explore this, we examined the temperature dependence of the Ce $f$-states and the Rh/Ge $s, p, d$ states from 50~K to 2000~K. We observed a strong temperature dependence across all states, characterized by the disappearance of the Fermi-level peaks at approximately 300~K (Fig.~\ref{fig2T:eDMFT_TDOS_PUC}(a, c, d)). Additionally, the temperature-dependent hybridization near $E_F$ (Fig.~\ref{fig2T:eDMFT_TDOS_PUC}(b)) is significantly stronger at low temperatures compared to high temperatures. Collectively, these features point to Kondo-like screening with a characteristic Kondo temperature ($T_K$) of approximately 300~K.
In Kondo systems, the local magnetic moment of the $f$-electrons is typically screened by conduction electrons at low temperatures. To determine the extent of this screening in Ce$_2$Rh$_3$Ge$_5$, we analyzed the local magnetic susceptibility ($\chi$) and the scattering rate ($\Sigma$) of the $j=5/2$ state in Fig.~\ref{fig2T:eDMFT_TDOS_PUC}. The susceptibility decreases continuously with falling temperature, showing a distinct change in slope around 300 K (Fig.~\ref{fig2T:eDMFT_TDOS_PUC}(e)), consistent with the suppression of the Kondo peak (Fig.~\ref{fig2T:eDMFT_TDOS_PUC}(a)). Similarly, the scattering rate is high at elevated temperatures but drops rapidly below 300~K (Fig.~\ref{fig2T:eDMFT_TDOS_PUC}(f)).
Due to the computational limitations of the method at low temperatures, we did not perform calculations below 50~K. Consequently, the susceptibility and scattering rate data cannot directly confirm the ground-state behavior of the Ce $f$-electron magnetic moment. However, if the local moment were completely screened, the scattering rate $\Sigma$ versus $T^2$ should extrapolate linearly to zero. In Fig.~\ref{fig2T:eDMFT_TDOS_PUC}(f), we plot $\Sigma$ vs. $T^2$ and applied linear fits to the regions below 150~K and 250~K (noting that the data deviates from linearity above 250~K). Both fits extrapolate to a finite scattering rate at $T=0$, implying that the local moment is not completely screened at very low temperatures.
In summary, our DFT+eDMFT calculations demonstrate the formation of a Kondo lattice in Ce$_2$Rh$_3$Ge$_5$ with a characteristic temperature of ~300 K, where the $s, p,$ and $d$ states of Rh and Ge contribute to the screening process. We also find indirect evidence that the Ce $f$-electron local moment remains partially unscreened at low temperatures, leaving room for magnetic interactions via the RKKY mechanism. To further clarify the low-temperature physics and anisotropic properties, we are currently employing a low-energy model assuming a localized $f$-electron screened by the surrounding Rh and Ge valence states.

 \subsection*{Low-energy theoretical model for hard-plane order}

 \begin{figure*}
  \centering
		\includegraphics[width=14.5 cm, keepaspectratio]{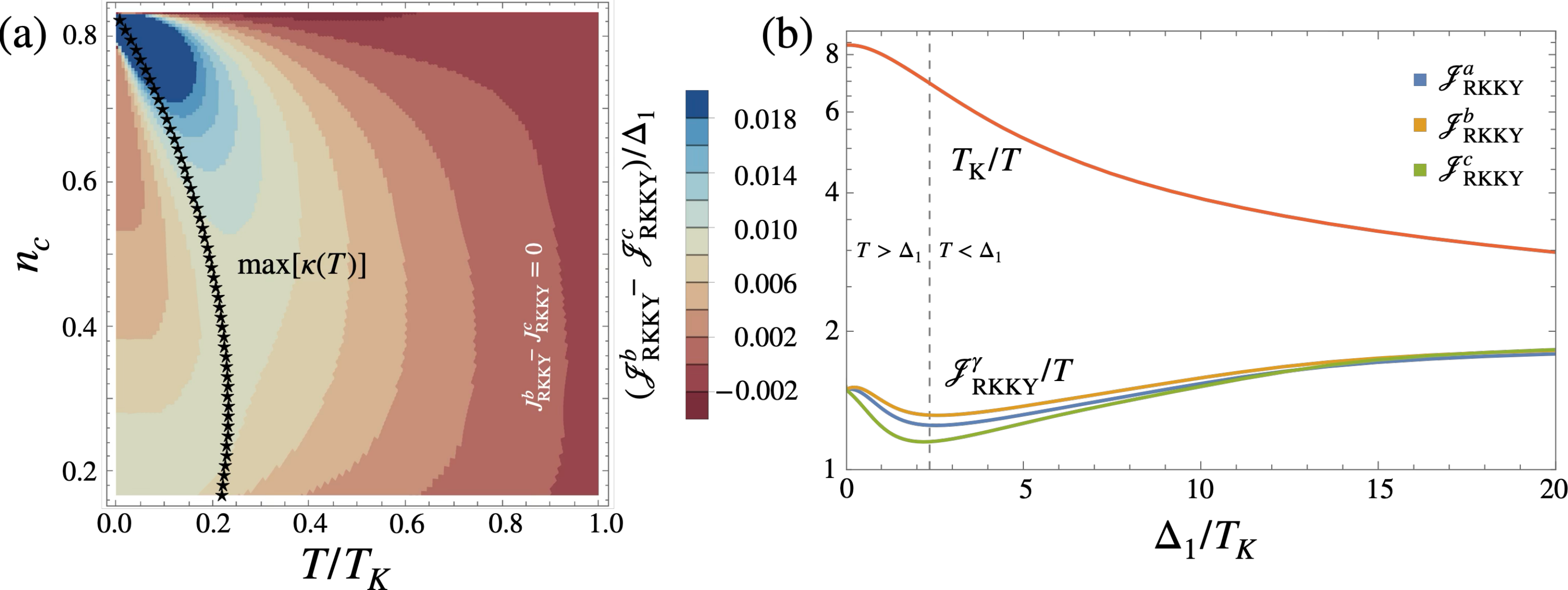}
		\caption{\textbf{Outcomes of the microscopic framework:} (a)  Difference in the strength of the RKKY exchange between uniformly $b$-polarised and $c$-polarised moments for the Coqblin-Schrieffer Hamiltonian in Eq.~(\ref{TM1}) in a constant density of states approximation, as a function of the electron filling $n_c$ and temperature and for a moderate Kondo coupling of $J_K/W=1/3$, where $W$ is the bandwidth, and in the weak CEF anisotropy limit $\Delta_{1,2}\rightarrow0$. At low temperatures the CEF easy-axis ($c$-axis) has a weaker uniform RKKY exchange than the CEF hard-axis ($b$-axis). This is congruous with the experimentally derived values of the molecular fields for Ce$_2$Rh$_3$Ge$_5$ where the AFM RKKY exchange is stronger in the hard $ab$-plane than along the CEF easy $c$-axis.  The black stars trace out the maximum of the $\kappa(T)$ function where the coexistence of local moments and heavy Fermi liquid is greatest and aligns well with the region with maximal splitting in the RKKY exchange along the $b$ and $c$ axes. (b) RKKY exchange for $f$-moments uniformly polarized along the three spin axes as a function of the anisotropy strength. The results refer to a moderate Kondo coupling $J_K/W=1/3$, band filling $n_c=0.75$, and a fixed temperature of $T/W=0.05$. The $ab$-plane has the strongest RKKY exchange up to very high anisotropy strengths of $\Delta_1/T_K\approx 12$. Increasing anisotropy also brings the RKKY and Kondo energy scales closer, further favoring the coexistence of Kondo hybridization and magnetic order.}
        \label{fig78:PD}
\end{figure*}

To shed light on the direction adopted by magnetic moments in $f$-electron systems under the influence of competing energy scales, we further employ low-energy models, using the Ce$_2$Rh$_3$Ge$_5$ system as an example.
Thus, below we discuss our low-energy model, which proposes an explanation for the hard-plane magnetic ordering in Ce$_2$Rh$_3$Ge$_5$.

The preceding eDMFT study, performed in the absence of any crystal electric fields, reveals a Kondo temperature of around $300$\,K, which in fact exceeds the CEF scale. Of course, the addition of CEFs will reduce the Kondo temperature. For moderate Kondo couplings and conduction band fillings, we find this to be at most a factor of 4 (see Fig.~\ref{fig78:PD}(b)), which is consistent with the estimate of Ref.~\cite{PhysRevB.60.10383} of $56$\,K. Although the coherence temperature is only slightly above the N\'eel temperature ($T^{\ast}_\mathrm{coh}=7.5$~K from the coherence maximum in the resistivity \cite{PhysRevB.60.10383}), the single-ion Kondo temperature $T_K$ is usually significantly higher. 
 Further, the presence of a two-stage Kondo effect in Ce$_2$Rh$_3$Ge$_5$ is evident from the logarithmic rise in the resistivity in the range $80-300$~K and another one in the range $7.5-30$~K below the  CEF energy scale $\Delta_2$. The ratio of the logarithmic slopes suggests that all six CEF states participate in Kondo scattering at high temperature, and the lowest two doublets participate down to $T^{\ast}_\mathrm{coh} \ll \Delta_1$. The fact that the lowest two doublets participate in Kondo scattering at temperatures 20 times smaller than $\Delta_1$ points to strong mixing of the CEF ground state with the first excited state and particle-hole excitations, and is perhaps responsible for the smaller gapping out of this excitation at lower frequencies as temperature is lowered. This dressing of the  ground state doublet  leads to an enhancement of the Kondo temperature: $T_{K}= T_{0K} \left(\frac{W}{2\Delta_1}\right)^2$ \cite{Coleman2015}, assuming for simplicity that all six CEF states participate to $T\sim \Delta_1$ and only the ground state doublet at lower temperatures ($W$ is the bandwidth and $T_{0K}$ is the Kondo temperature for a single doublet). The strong Kondo hybridization that precedes N\'eel order is further evident from the fact that the ordered moments ($\mu_x$ = 0.24(1)~$\mu_{\text{B}}$, $\mu_y$ = 0.19(1)~$\mu_{\text{B}}$) are significantly smaller than the saturation moments for the ground state doublet ($\mu_x=0.78\mu_{\text{B}}$, $\mu_y=0.64\mu_{\text{B}}$).

 Previous works have shown how the Kondo interaction can generate an exchange anisotropy that opposes the CEF anisotropy \cite{VojtaBrando, Scott2025, Hanshang}. This can lead to magnetic order that is perpendicular to the CEF easy-axis, and the phenomenon is indeed observed in many heavy-fermion compounds \cite{Brando2019}. For Ce$_2$Rh$_3$Ge$_5$, the CEF easy-axis is the $c$-axis. Yet, it orders in the hard $ab$-plane. The molecular fields $0<-\lambda_{c} \ll-\lambda_{a,b}$ suggest that the RKKY interaction is AFM and is much stronger in the hard $ab$-plane, which is most likely the reason for the hard-plane AFM order below the N\'eel temperature.

 Ref.~\cite{Scott2025} shows how the development of coherence and partial Kondo hybridization of the local moments with conduction electrons  generates a stronger RKKY exchange along the hard CEF direction that can result in magnetic order along this direction. Here, starting with the Coqblin-Schrieffer model of the Ce $4f^1$ moment subject to the CEF scheme derived in Eq.~(\ref{H-CEF}), we show how Kondo hybridization leads to a stronger RKKY exchange in the $ab$-plane within a simple constant DOS approximation for the conduction band. 
 
 The Coqblin-Schrieffer Hamiltonian with general anisotropy is given as
 \begin{eqnarray}
     H_\mathrm{CS} &=& \sum_{i\alpha\beta}D_{\alpha \beta} f^{\dagger}_{\alpha i}f_{\beta i}+\sum_{\mathbf{k}\alpha}^{|\epsilon_\mathbf{k}|\leq W/2}\epsilon_{\mathbf{k}} c^{\dagger}_{\alpha\mathbf{k}}c_{\alpha\mathbf{k}}
     \nonumber\\
     &&-J_K \sum_{i\alpha \beta} f^{\dagger}_{\alpha i}  c^{\dagger}_{\beta i} f_{\beta i}c_{\alpha i},
     \label{TM1}
 \end{eqnarray}
where $f_{\alpha i}$ ($f^\dagger_{\alpha i}$) annihilates (creates) an impurity fermion with angular momentum $\alpha\in\{-5/2,\,-3/2,\,\dots,\,3/2,\,5/2\}$ on site $i$ and $c_{\alpha i}$ ($c^\dagger_{\alpha i}$) annihilates (creates) a conduction fermion with angular momentum $\alpha$ on site $i$. $D_{\alpha\beta}$ are linear combinations of the Stevens parameters defined in Eq.~(\ref{H-CEF}), $W$ is the bandwidth and $J_K$ is the Kondo coupling. Motivated by the eDMFT results, we will consider a moderately strong Kondo coupling, where Kondo-driven hard-direction order is possible, and will consider a representative value of $J_K/W=1/3$ throughout. (At weak enough Kondo coupling, there is no Kondo hybridization and the RKKY exchange obtained from the model in Eq. \ref{TM1} is isotropic.)

In the mean-field approximation, the Coqblin-Schrieffer Hamiltonian becomes
\begin{eqnarray}
    H_\mathrm{MF}&=&\sum_{i\alpha\beta}D_{\alpha \beta} f^{\dagger}_{\alpha i}f_{\beta i}+\sum_{\mathbf{k}\alpha}\epsilon_{\mathbf{k}} c^{\dagger}_{\alpha\mathbf{k}}c_{\alpha\mathbf{k}}
     \nonumber\\
     &&-J_K V \left(\sum_{i} f^{\dagger}_{\alpha i}  c^{}_{\alpha i} +{\rm h.c.}\right)+\lambda\sum_i f^\dagger_{\alpha i}f_{\alpha i},\\
     V&=&\sum_{\alpha}\langle f^\dagger_{\alpha i}c_{\alpha i}\rangle,
      \label{TM2}
\end{eqnarray}
where $V$ is the hybridization order parameter and $\lambda$ a Lagrange multiplier that ensures there is one $f$-electron per site.  

Unfortunately, the mean-field Coqblin-Schrieffer Hamiltonian is not amenable to capturing a phase of coexistent magnetic order and Kondo hybridization. However, even in the Kondo paramagnet phase, we can  probe the anisotropy of the exchange between local $f$-moments. This can be done in several ways. For example, we can find the susceptibility of the Kondo paramagnet in the temperature window where the $f$-electrons still retain some local-moment character and extract the molecular fields. We obtained similar results using this approach to the one we propose now. We relocalise an infinitesimal fraction of the $f$-electrons and integrate out the remaining electrons to obtain the following RKKY interaction
\begin{eqnarray}
    H_\mathrm{RKKY}=-\frac{1}{2}\sum_{i,j}\mathcal{J}_\mathrm{RKKY}^{\gamma}(\mathbf{r}_i-\mathbf{r}_j) S_i^{\gamma}S_j^{\gamma},
     \label{TM3}
\end{eqnarray}
where $S^{\gamma}_i$ is the relocalized local moment on site $i$ aligned along direction $\gamma$. Note that, for each lattice site, the constraint on the $f$-electron number becomes $\sum_{\alpha}f^{\dagger}_{\alpha i}f_{\alpha i} =1\rightarrow \sum_{\alpha}\left(f^{\dagger }_{\alpha i}f_{\alpha i}+F^{\dagger}_{\alpha i}F_{\alpha i} \right) =1$, where $F_{\alpha i}$ $(f_{\alpha i})$ denote localized (delocalized) $f$-electrons. As in the usual mean-field approximation, the constraint is softened and enforced only globally with the localized $f$-electrons taking up an infinitesimal fraction. We have treated the relocalized moments classically and for a moment aligned along the angular momentum quantisation axis we have taken the relocalized fermionic operators to have an expectation value of $\langle F^{\dagger}_{\alpha}F_{\alpha} \rangle = \left(\frac{1}{7}+\frac{2}{35} \alpha \right) S^z $, with $S^z\rightarrow 0$. With this, the anisotropic RKKY exchange becomes
\begin{eqnarray}
    &&\mathcal{J}_\mathrm{RKKY}^{\gamma}(\mathbf{r}_i-\mathbf{r}_j) \nonumber\\
    &&=\left(\frac{2}{35}\right)^2J_K^2\sum_{\alpha\beta}\alpha\beta \int_0^{\beta} d \tau 
    \langle c^{\dagger}_{\alpha i}(\tau) c_{\beta j} (0)\rangle
     \langle c_{\alpha i}(\tau) c_{\beta j}^{\dagger} (0)\rangle,
      \label{TM4}
      \nonumber\\
\end{eqnarray}
where the expectation values are taken with respect to a rotated mean-field Hamiltonian such that $\gamma$ is the angular momentum quantisation axis. Note that we have not included the {\it isotropic} energy penalty associated with relocalising a fraction of the $f$-moments. This energy cost is the reason why there is a vanishing local moment as $T\rightarrow 0$ within the mean-field description of the Coqblin-Schrieffer model at moderate Kondo couplings.   

Ref. \cite{Scott2025} found that for a spin-1 underscreened Kondo lattice model with sufficiently strong Kondo coupling, the order is along the CEF hard-axis due to stronger RKKY exchange, irrespective of whether it is FM or AFM order.  Motivated by this, for the sake of simplicity, we restrict our attention to models where FM order is preferred and the band of bandwidth $2W$ can be approximated by a constant DOS. We calculate the RKKY exchange for the zero wavevector $\mathcal{J_\mathrm{RKKY}(\mathbf{q}=\mathbf{0}})$ (we have observed qualitatively similar RKKY anisotropy for the Brillouin zone average). The results for the zero wavevector are shown in Fig.~\ref{fig78:PD}. The figure shows the splitting in the $\mathbf{q}=\mathbf{0}$ RKKY exchange along the $b$ and $c$ spin axes in the limit of the anisotropy $\Delta_{1,2}\rightarrow 0$, as a function of conduction band filling and temperature. $n_c$ is the number of conduction electrons interacting with the impurity per site per band.
Kondo hybridization leads to the CEF hard-axis acquiring a stronger RKKY exchange at sufficiently low temperatures.
Similar to what was observed in an underscreened Kondo lattice model~\cite{Scott2025}, the insulator ($n_c=5/6$ in the present model) does not have an RKKY exchange that opposes the CEF anisotropy. 

The splitting in the exchange appears to be maximised when the system is deep inside the "two-fluid" phase and the coexistence of local moments and the heavy-fermion liquid is maximised. This happens when the coherent heavy Fermi liquid begins to develop but the magnetic moment of the $f$-electrons is still a sizable fraction of its full free-spin value. To capture the two-fluid phase we define the parameter, $\kappa$, as such
\begin{eqnarray}
    \kappa(T)&=&\frac{\chi_\mathrm{cf}(T)}{\chi_\mathrm{cf}(0)}\frac{T\chi_\mathrm{ff}(T)}{\frac{j(j+1)}{3}\left( 1-\frac{1}{2j+1}\right)}
     \label{TM5},
\end{eqnarray}
where $\chi_\mathrm{ff}$ ($\chi_\mathrm{cf}$) is the susceptibility of the $f$- (conduction) fermions when a magnetic field is applied only to the $f$-fermions and $\frac{j(j+1)}{3T}\left( 1-\frac{1}{2j+1}\right)$ is the free-spin susceptibility of the $f$-moment (with a $1/N$ correction arising from the fact that the single occupancy constraint is only enforced on average at the mean-field level). As in Ref.~\cite{PhysRevB.70.235117}, we use the magnitude of $\chi_\mathrm{cf}(T)$ as the measure of the coherent heavy Fermi liquid component which we define as $\chi_\mathrm{cf}(T)$/$\chi_\mathrm{cf}(0)$ since $\chi_\mathrm{cf}(T)$ approaches its maximum as $T\rightarrow 0$ and there is no local-moment component left. We take $T\chi_\mathrm{ff}$ as a measure of the local-moment component since it interpolates between $\frac{j(j+1)}{3}\left( 1-\frac{1}{2j+1}\right)$ for a free $f$-moment at high temperature and $0$ as $T\rightarrow0$ and the local moment vanishes. Thus, the coexistence of the local-moment and Fermi liquid components is strongest when $\kappa$ is maximized. Fig.~\ref{fig78:PD}(a) shows how $\mathrm{max}\left[\kappa(T)\right]$ closely follows the maximum splitting in the RKKY exchange along the b and c spin axes. In agreement with Ref. \cite{Scott2025}, we can thus see that the anisotropy in the RKKY exchange, which opposes the anisotropy of the CEF field, is a consequence of the coexistence of the local moments and the heavy-fermion liquid. This can then give a lower energy if the moments polarise along the hard axis.

We can understand the stabilization of ordering along the hard-axis by considering a weak polarisation of the conduction electrons. The $f$-electrons that are tied to the residual local moment antialign locally with the conduction electrons, whereas those that delocalise and are part of the Fermi liquid quasiparticle align locally with the conduction electrons ($\chi_{cf}(0)>0$). This frustration is relieved by CEF anisotropy, which lowers (increases) the energy of the coexistent two-fluid state along the hard (easy) axis. Because the mechanism relies on the {\it local} alignment of $f$-moments and conduction electrons, it is perhaps not surprising that it can lead to hard-direction order that is antiferromagnetic or ferromagnetic. The wavevector $\mathbf{q}$ with which the magnetic order is modulated will depend on the details of the band structure and the relative positions of the magnetic impurities, but the hard direction of the order results from the local interplay of Kondo hybridisation, residual moment polarisation and CEF effects. In fact, recent NRG calculations for a single-impurity $j=3/2$ Coqblin-Schrieffer model have shown that in the temperature window where the impurity is partially screened, the susceptibility of the impurity dressed with the Kondo cloud is stronger along the CEF-hard direction, provided that the Kondo coupling is strong enough \cite{wojcik2026}. There, the frustration coming from the coexistence of partially screened, polarised local moment and Kondo hybrdization manifested through the fact that $\langle S_{f}S_c\rangle<0$ and $\langle S_c \rangle,\langle S_f \rangle >0$, where $S_f$ and $S_c$ are the magnetisations of the impurity and conduction electrons respectively. These reveal that for the partially delocalized moment the entangled $f$ and $c$ electrons polarise the same way, while maintaining local antialignment.   Extrapolating the single-impurity calculation to the lattice, again one would expect the partially screened impurities, dressed with Kondo clouds, to prefer magnetic order along the hard-axis.

Fig.~\ref{fig78:PD}(b) shows the splitting in the uniform RKKY exchange as a function of anisotropy at a fixed temperature $T/W=0.05$ and Kondo coupling strength $J_K/W=1/3$. We can see that the splitting grows with anisotropy before becoming smaller again. However, the exchange in the $ab$-plane is strongest all the way to $\Delta_1/T_K \approx 12$. In fact, for the CEF parameters of Ref. \cite{UMEO2002403}, this is the case all the way to arbitrarily large anisotropies (see Supplementary Fig.~2).
Increasing anisotropy brings the RKKY and Kondo energy scales closer, thus favoring the coexistence of magnetic order and Kondo hybridization.

The results of the low-energy Coqblin-Schrieffer model align well with the conclusions reached by the high-temperature eDMFT study. For example, Fig.~\ref{fig78:PD}(a) shows that the splitting in the RKKY exchange increases with the number of screening conduction electrons ($n_c$), as a result of a stronger Kondo effect. eDMFT does indeed point to a significant hybridisation of the $f$-moment with both the $p$ and the $d$ orbitals of all Rh and Ge atoms, and hence likely a large number of screening conduction electrons (see Figs. \ref{fig1T:eDMFT_DOS_PUC}(c) and \ref{fig2T:eDMFT_TDOS_PUC}(a)-(d)). Further, eDMFT shows a Kondo temperature scale of around 300\,K in the absence of anisotropy. This is consistent with the estimate of Ref. \cite{PhysRevB.60.10383} of 56\,K in the presence of anisotropy, and the Coqblin-Schrieffer model predicting a particularly strong enhancement of RKKY exchange for the hard-plane at the corresponding value of $\Delta_1/T_K \sim 193/56 \sim 3.4 $ (see Fig. \ref{fig78:PD}(b)). Finally, the extrapolation of the eDMFT data to low temperatures suggests that the $f$-moments are partially screened in this regime (see Fig. \ref{fig2T:eDMFT_TDOS_PUC}(f)). This aligns well with the prediction of the low-energy model, which shows that the strongest enhancement of the RKKY exchange in the hard-plane occurs in the low-temperature regime where the coexistence of partially screened local moments and heavy-fermion liquid is greatest.

 \section*{Discussion}

Ce$_2$Rh$_3$Ge$_5$ provides a rare realization of a Kondo lattice in which the static ordered moments lie in a CEF-hard plane. INS and susceptibility yield a ground-state Kramers doublet dominated by the $\lvert \mp 3/2\rangle$ component of the $J=5/2$ multiplet, producing a larger effective $g$ factor for $H\parallel c$ and thus a single-ion easy $c$ axis in the paramagnetic regime. In contrast, NPD shows that the ordered Ce moments lie within the $ab$ plane at $T = 1.6$~K, forming a commensurate AFM structure with a substantially reduced ordered moment, $\mu_{\rm Ce}\approx0.31\,\mu_{\rm B}$. The ZF-$\mu$SR spectra, which display a single well-defined precession frequency corresponding to $B_{\rm loc}\simeq25$\,mT at the dominant muon site, confirm that this hard-plane AFM state is homogeneous and bulk.

Hard-axis and hard-plane ordering have been reported primarily in Ce- and Yb-based {\it ferromagnetic} Kondo lattices such as CeAgSb$_2$~\cite{PhysRevB.104.115169}, YbNi$_4$P$_2$~\cite{Krellner_2011}, and Ce$_2$PdGe$_3$~\cite{PhysRevB.94.014418,PhysRevB.91.035102}. In these systems, the ordered moments lie perpendicular to the high-$T$ CEF easy direction inferred from $\chi(T)$, often attributed to strongly anisotropic RKKY interactions associated with low-symmetry coordination environments or multiple Ce sites~\cite{Hanshang}. Ce$_2$PdGe$_3$ illustrates this interplay vividly: an AFM transition along $c$ is followed by a basal-plane ferromagnetic transition at lower $T$, yielding a hard-plane FM state. Our results extend this phenomenology in three important ways: (i) the ordering in Ce$_2$Rh$_3$Ge$_5$ is AFM, rather than FM; (ii) the hard-plane state appears already at the primary magnetic transition, without a secondary reorientation; and (iii) the Kondo renormalization is substantially stronger, as reflected in the small ordered moment and in the fact that $T_{\mathrm{N}}$ is much smaller than the estimated Kondo scale $T_K$.

From the CEF fits we obtain molecular field constants $\lambda_{a}\simeq -54$, $\lambda_{b}\simeq -47$, and $\lambda_{c}\simeq -6.6$\,mol/emu, implying AFM RKKY exchange along all three directions, but nearly an order of magnitude stronger in-plane than along $c$. Such exchange anisotropy is compatible with what is required to overcome the single-ion CEF preference for $c$-axis polarization and stabilize a hard-plane AFM state. In this respect Ce$_2$Rh$_3$Ge$_5$ is consistent with the conclusion drawn from recent surveys of Ce-based Kondo-lattice ferromagnets that hard-axis or hard-plane order is primarily governed by anisotropic RKKY interactions~\cite{Hanshang}.

Such a mechanism has been proposed in theoretical studies of itinerant magnets, where transverse quantum fluctuations near a ferromagnetic quantum critical point can favor magnetization along a CEF-hard direction~\cite{PhysRevLett.113.147001}. Extensions of this idea to Kondo lattices suggest that Kondo hybridization leading to a partial moment delocalization can similarly renormalize the exchange such that RKKY interactions are strongest along a CEF-hard direction. Our theoretical analysis of Ce$_2$Rh$_3$Ge$_5$, based on an anisotropic Coqblin--Schrieffer model parametrized using the experimentally determined CEF scheme, provides a concrete realization of this scenario. By integrating out the heavy quasiparticles in a mean-field treatment, we obtain an effective RKKY Hamiltonian of the form given in Eq.~\ref{TM3}. For moderate Kondo coupling $J_K/W\approx1/3$ and conduction-band filling $n_c\approx0.75$, the calculated uniform exchange $\mathcal{J}^{\gamma}_{\rm RKKY}(\mathbf{q}=0)$ is largest for spins polarized in the $ab$ plane over a wide range of CEF anisotropies, with $\mathcal{J}_{\rm RKKY}^{ab}-\mathcal{J}_{\rm RKKY}^{c}>0$ up to $\Delta_1/T_K\sim10$. The splitting between the in-plane and out-of-plane exchange is maximal in the temperature window where the two-fluid parameter (Eq.~\ref{TM5}) peaks, signalling the strongest coexistence of local-moment and heavy-fermion components. This suggests a microscopic link between Kondo hybridization and the anisotropic RKKY exchange that stabilizes hard-plane order: the same hybridization that reduces the static ordered moment also enhances the in-plane exchange relative to the $c$ direction.

Our findings therefore establish Ce$_2$Rh$_3$Ge$_5$ as a system lying at the intersection of two previously distinct lines of thought. On one hand, the strong in-plane molecular fields and the comparison with other Ce compounds support the view that hard-axis/plane order is primarily governed by anisotropic RKKY interactions shaped by the lattice geometry and electronic structure. On the other hand, the pronounced Kondo renormalization and the success of our Coqblin--Schrieffer analysis demonstrate that these anisotropic exchanges can themselves be generated and amplified by Kondo-driven two-fluid physics. Ce$_2$Rh$_3$Ge$_5$ therefore shows that a single AFM transition in a strongly hybridized Kondo lattice is sufficient to realize a hard-plane state, without requiring ferromagnetism or multiple magnetic transitions. Future pressure or chemical-substitution studies, which tune the balance between $T_K$ and $T_{\mathrm{N}}$, should provide a stringent test of the predicted correlation between the two-fluid regime and the magnitude of the hard-plane exchange anisotropy.

Thus, Ce$_2$Rh$_3$Ge$_5$ realizes a rare AFM Kondo lattice in which the static ordered moments reside in a CEF hard-plane. NPD establishes a commensurate in-plane AFM structure with a strongly reduced ordered moment, while ZF-$\mu$SR confirms the emergence of homogeneous bulk magnetic order below $T_{\mathrm{N}}^{\mu}\approx 5.34$\,K. Complementary INS measurements reveal a well-defined orthorhombic CEF level scheme and a ground-state Kramers doublet that favors $c$-axis polarization in the single-ion limit. The contrasting orientations selected by the CEF and by the collective ordered state demonstrate that anisotropic RKKY exchange overwhelms single-ion anisotropy at low temperatures in the presence of substantial Kondo renormalization. Our theoretical analysis, based on an anisotropic Coqblin--Schrieffer model with the experimental CEF parameters as input, identifies a microscopic mechanism for this inversion: partial Kondo hybridization amplifies the RKKY exchange along the CEF-hard direction in the two-fluid regime, precisely where the coexistence of local moments and heavy quasiparticles is strongest.

Taken together, these results establish Ce$_2$Rh$_3$Ge$_5$ as a model system for hard-plane magnetic ordering in a Kondo lattice AFM. The combination of precise magnetic structure determination, microscopic confirmation of bulk order, and quantitative knowledge of the CEF scheme provides a stringent experimental foundation for theories in which Kondo hybridization generates anisotropic exchange that competes with the CEF. The present work therefore not only clarifies the microscopic origin of the hard-plane state in Ce$_2$Rh$_3$Ge$_5$, but also demonstrates how the balance of Kondo and RKKY interactions can stabilize magnetic order along a CEF hard direction in heavy-fermion materials.

\section*{METHODS}
\label{expt}

\subsection*{Experimental details}

Polycrystalline samples of Ce$_2$Rh$_3$Ge$_5$ and its nonmagnetic analog La$_2$Rh$_3$Ge$_5$ were synthesized by arc melting stoichiometric amounts of the constituent elements Ce (3N purity), Rh (3N), Ge (5N), and La (3N) on a water-cooled copper hearth under an argon atmosphere. The La-based compound was synthesized specifically to estimate the phonon contribution in Ce$_2$Rh$_3$Ge$_5$. The arc-melted buttons were subsequently sealed in evacuated quartz ampoules and annealed at 1100$^\circ$C for 1 day, followed by 1000$^\circ$C for 7 days to ensure homogeneity. Phase purity and structural characterization were performed at room temperature using powder x-ray diffraction (XRD) with Cu K$_{\alpha}$ radiation on a commercial diffractometer.

NPD measurements were performed on a 8-g sample of Ce$_2$Rh$_3$Ge$_5$ on the high intensity two-axis D20 diffractometer at the Institut Laue Langevin (ILL), France, in the high-flux mode, with an incident wavelength of 2.42\,\AA\ at 1.6, 4, and 10\,K. The high-resolution powder diffractometer D2B (ILL) was used, with a wavelength of 1.594\,\AA\ for the structural investigation at room temperature.

INS measurements were conducted on the High-Resolution Chopper Spectrometer (HRC) at the BL12 beamline of the Materials and Life Science Experimental Facility (MLF), Japan Proton Accelerator Research Complex (J-PARC)~\cite{ITOH201190}. Three incident energies were employed using a monochromatic Fermi chopper: $E_i = 10$~meV at 100\,Hz, to probe the low-energy magnetic excitations below and near $T_{\text{N}}$, and $E_i = 60$ and 150~meV at 300\,Hz, to access the higher-energy CEF excitations. The experiments were performed over a temperature range of 4--300~K using a bottom-loading closed-cycle refrigerator. Measurements on the nonmagnetic analog La$_2$Rh$_3$Ge$_5$ were performed under similar conditions to estimate and subtract the phonon background. Further details about the spectrometer configuration and resolution can be found in Ref.~\cite{ITOH201190}. Additionally, a brief confirmatory measurement was performed at 15~K on the MERLIN time-of-flight spectrometer at the ISIS Facility (Supplementary Fig.~1(b))~\cite{BEWLEY20061029}, 
primarily to obtain a larger $Q$-coverage.

Zero-field (ZF) $\mu$SR experiments were performed at the VMS spectrometer of the Swiss Muon Source (S$\mu$S), Paul Scherrer Institute (PSI), Villigen, Switzerland. We used a 13-mm diameter $\times$ 2-mm thick powder pellet, which was mounted on the sample stick of a He-4 cryostat. The temperature range of the measurements spanned from 1.8 to 5.4\,K, allowing us to investigate the magnetic ordering transition and the low-temperature magnetic ground state.

\subsection*{Theoretical details}

In this paper, calculations are performed using a state-of-the-art approach based on density functional theory combined with embedded dynamical mean-field theory (DFT+eDMFT) for correlated materials, as implemented in the eDMFT code \cite{eDMFT1,eDMFT2,eDMFT3}. The WIEN2k code was used for the DFT component, which is based on the full-potential linearized augmented plane-wave (LAPW) plus local orbitals method ~\cite{wien2k}. A unique aspect of DFT+eDMFT is its use of quasi-atomic orbitals—which are more localized than Wannier orbitals—to describe the interacting part of the Hamiltonian, while the single-particle part is captured within the complete LAPW basis ~\cite{eDMFT1,eDMFT2}. The radial part of quasi-atomic orbitals is the solution of the Schr\"odinger equation in the muffin-tin sphere, with linearized energy at the Fermi level, and the angular dependence given by the spherical harmonics. The Perdew-Burke-Ernzerhof generalized gradient approximation (PBE-GGA) is employed for the exchange-correlation potential \cite{PBE}. Ce $4f$ electrons are treated dynamically with the DMFT local self-energy, where the full atomic interaction matrix is taken into account with a nominal double-counting  \cite{nominalDC}. The DMFT calculations consider only the local self-energy of the $4f$ orbital, while the other orbitals are treated at the DFT level. The local correlation effect is contained in the self-energy $\Sigma(\omega)$, which is calculated from the impurity problem. To solve the impurity problem, we used CTQMC (Continuous-Time Quantum Monte Carlo) \cite{CTQMC,CTQMC1}. We used a large hybridization window of $\pm$ 10 eV. For the Monte Carlo (MC) runs we employed at least 200 x $10^6$ MC steps.
Strong spin-orbit coupling causes the $4f$ states to split into two manifolds, corresponding to the total angular momentum quantum numbers $j=5/2$ and $j=7/2$. We used interaction parameters, $U=6.0 eV$ and $J=0.7eV$, similar to previous works \cite{eDMFT1,fBM}. The occupation of the Ce $4f$ orbital was estimated to be 0.97 at all temperatures. We neglected the crystal electric field (CEF) in the local correlation energy because it has been experimentally confirmed as a $j=5/2$ system and we aim to obtain the isotropic properties of the system. Recent studies have demonstrated that DFT+eDMFT is predictive of both the electronic and structural properties of correlated materials at finite temperatures. This predictive capability has also been showcased in other $f$-electron systems \cite{fBM,fBM1}. The local magnetic susceptibility is computed via quantum impurity solver sampling.

\section*{Data Availability Statement}
Data are available from the corresponding author upon reasonable request.

\section*{Code Availability}

The DFT calculations were performed using the WIEN2k code, and embeded DMFT calculations with the open-source eDMFT code developed by Kristjan Haule at Rutgers University (http://hauleweb.rutgers.edu/tutorials/). Haule’s
eDMFT code is freely distributed for academic use under the Massachusetts Institute of Technology (MIT) licence.

\begin{acknowledgments}
We gratefully acknowledge the MLF/J-PARC facility for the allocation of beam time on the HRC spectrometer (Proposal No:2022A0359), the ISIS Facility for beam time on MERLIN (RB2490291), ILL on the D20/D2B diffractometers (the beam time allocation under the experiment code 5-31-2920, doi:10.5291/ILL-DATA.5-31-2920), and PSI on the S$\mu$S VMS spectrometer. RT thanks the SERB for the National Post Doctoral Fellowship (PDF/2022/001539). DTA would like to thank the Royal Society of London for International Exchange funding between the UK and Japan, Newton Advanced Fellowship funding between UK and China and EPSRC UK for the funding (Grant No.\ EP/W00562X/1). DTA thanks CAS for PIFI Fellowship. We thank K.\ Umeo for providing us with the magnetization data at 2~K and Leandro Liborio for allowing to use muon site calculations.
Computing resources for the DFT+eDMFT theoretical calculations were provided by STFC Scientific Computing Department’s SCARF cluster. DFT+eDMFT work is funded by European Commission within the framework of the Romanian National Recovery and Resilience Plan (PNRR-III-C9-2022-I8 grant) through the ESCARGOT project entitled “Enhanced Single Crystal Applications and Research in the Growth of new Optical rare earth-based compounds for sustainable and efficient Technologies” (Code 136/15.11.2022 ; CF n°760080/23.05.2023).
\end{acknowledgments}

\section*{Author contributions}
D.T.A., R.T., and D.D. initiated the project. T.T. and T.O. synthesized the samples and performed XRD measurements. C.R. and D.T.A. carried out the neutron diffraction experiments, and C.R. analyzed the neutron diffraction data. D.D. and T.S. performed the $\mu$SR experiments, and R.T., D.D., F.P., and T.S. analyzed the $\mu$SR data with input from D.T.A.

R.T. and D.T.A. led the inelastic neutron scattering (INS) measurements and analysis, with experiments performed on HRC (J-PARC) by D.T.A., R.T., T.M., S.A., T.O., and S.L., and on MERLIN (ISIS) by D.T.A. and R.T. and D.T.A. carried out the primary INS data reduction, analysis, and interpretation.

M.K., E.S., and H.H. performed theoretical calculations addressing the hard-plane ordering. G.L.P. and N.C. performed density functional theory plus embedded dynamical mean-field theory (DFT+eDMFT) calculations. R.T., D.T.A., C.R., D.D., M.K., E.S., and G.L.P. drafted the manuscript and prepared the supplementary material, with additional input from S.L., T.T., O.T., T.S., A.S.M., A.S., and S.P. All authors discussed the results and contributed to the final manuscript.

\section*{Competing interests}
The authors declare no competing interests.

\section*{Supplementary materials}
\section{Low-energy INS and CEF analysis}
\setcounter{figure}{0}
\renewcommand{\thefigure}{A\arabic{figure}}

\begin{figure*}
		\includegraphics[width=14cm, keepaspectratio]{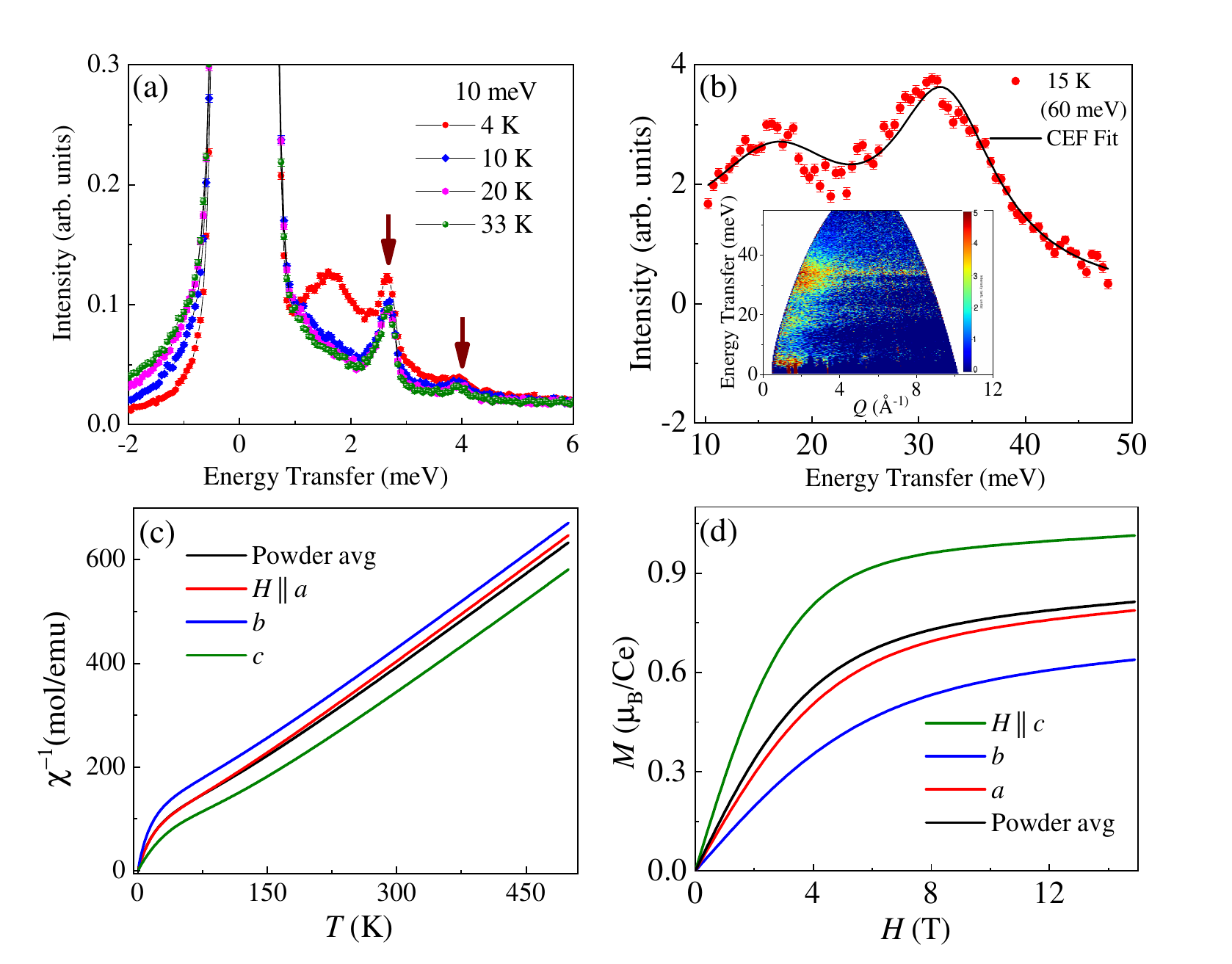}
		\caption{(a) Momentum-integrated ($Q = 0$–$1~\text{\AA}^{-1}$) low-energy INS spectra of Ce$_2$Rh$_3$Ge$_5$ measured on the HRC spectrometer using $E_i = 10$~meV at several temperatures across the N\'eel temperature. (b) One-dimensional energy cut of the magnetic INS spectrum measured on the MERLIN spectrometer with $E_i = 60$~meV (shown in the inset), demonstrating that the same CEF parameters obtained from the HRC measurements provide an equally good description of the higher-energy magnetic response. (c) CEF simulations of the inverse magnetic susceptibility $\chi^{-1}(T)$ and magnetization $M(H)$ along the principal crystallographic directions, calculated using the refined CEF parameters, reproducing the experimentally observed magnetic anisotropy.}
		\label{INS3}
\end{figure*}

Fig.~\ref{INS3}(a) shows the temperature dependence of the low-energy INS spectra of Ce$_2$Rh$_3$Ge$_5$ measured on the HRC spectrometer with $E_i = 10$~meV, after integrating over the low-$Q$ range $Q = 0$–$1~\text{\AA}^{-1}$. At 4~K, a clear excitation is observed near 1.7~meV, which rapidly weakens with increasing temperature and is absent above $T_{\mathrm{N}}$. This behavior identifies the 1.7~meV mode as a magnetic excitation associated with the antiferromagnetically ordered state. In contrast, additional features near 2.5~meV and 4~meV remain temperature independent and persist well above $T_{\mathrm{N}}$, indicating that they arise from spurious multiple scattering from the cryostat, as marked by arrows.

To further verify the robustness of the CEF scheme, an express INS measurement was performed on the MERLIN spectrometer using $E_i = 60$~meV. The magnetic spectra exhibit excitations near 16.6 and 32.3~meV, which are well reproduced using the same CEF parameters obtained from the HRC data, without additional fitting (Fig.~\ref{INS3}(b)).

Using these parameters, simulations of the inverse susceptibility $\chi^{-1}(T)$ and magnetization $M(H)$ along the principal crystallographic directions reproduce the experimentally observed anisotropy in the paramagnetic regime, confirming the $c$ axis as the CEF easy axis (Fig.~\ref{INS3}(c) and (d)). Combined with neutron diffraction results showing in-plane ordered moments, these consistency checks further substantiate the conclusion of hard-plane AFM ordering in Ce$_2$Rh$_3$Ge$_5$.

\section{Splitting in the RKKY interaction for the CEF parameters of Ref. \cite{UMEO2002403}}
\begin{figure}
    \centering
    \includegraphics[width=0.98\linewidth]{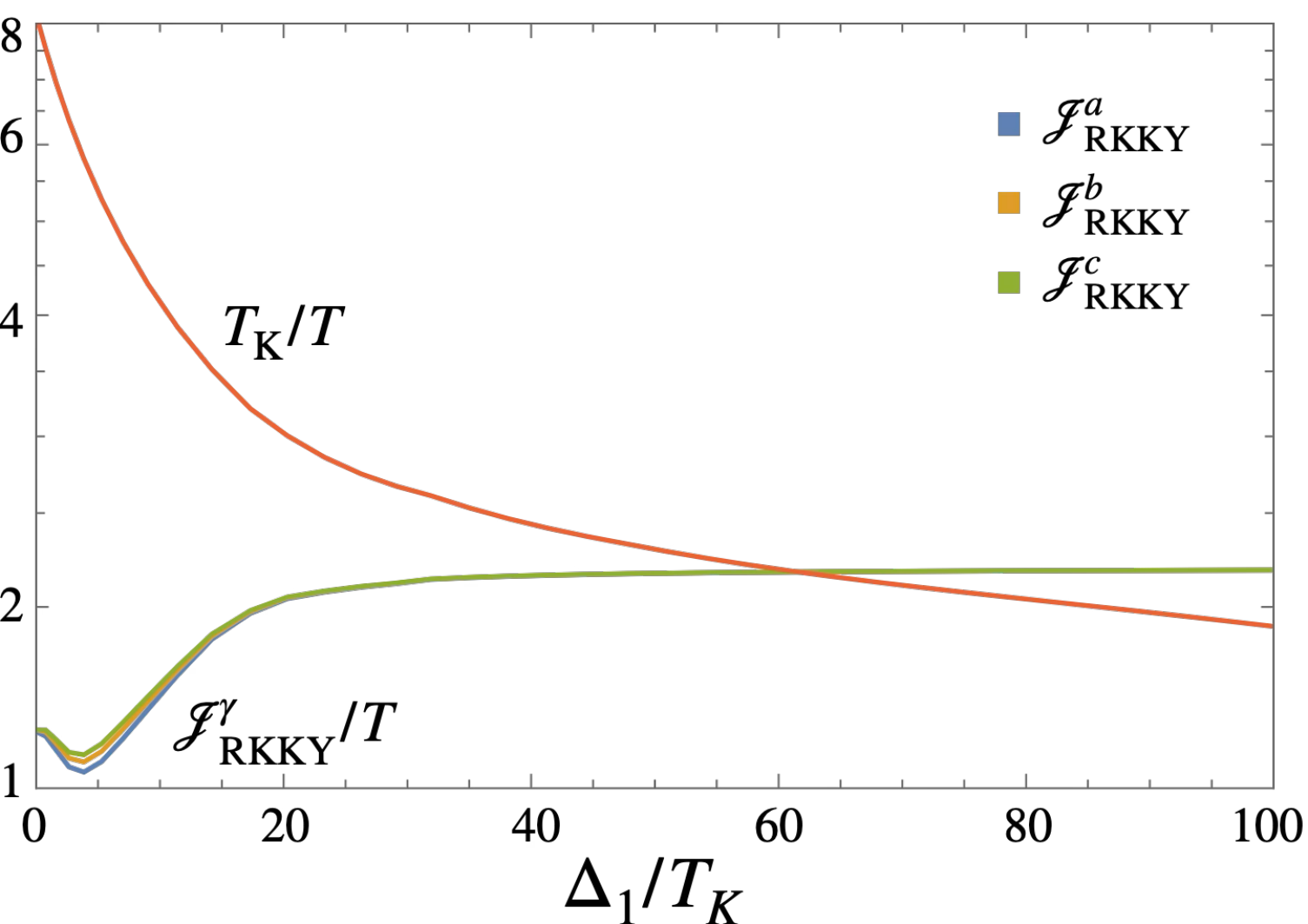}
    \caption{RKKY exchange for $f$-moments uniformly polarized along the three spin axes as a function of the anisotropy strength for the CEF parameters given in Ref. \cite{UMEO2002403}. The results refer to a moderate Kondo coupling $J_K/W=1/3$, band filling $n_c=0.75$, and a fixed temperature of $T/W=0.05$. The $ab$-plane has the strongest RKKY exchange up to arbitrary high anisotropy, with strengths of up to $\Delta_1/T_K> 100$ plotted.}
    \label{fig:oldpaperRKKY}
\end{figure}

Fig. \ref{fig:oldpaperRKKY} shows the splitting in the RKKY for the CEF parameters given in Ref. \cite{UMEO2002403}. There is no crossing of the different axes at large anisotropy and the b-axis remains the easy axis up to arbitrary large anisotropies. Additionally, the splitting between the a- and b- axis is always smaller than the splitting between either the a- or b- axis and the c-axis. This is consistent with the experiment as the moments lie within the ab plane for ${\rm Ce_2Rh_3Ge_5}$.

\FloatBarrier
\bibliography{bibliography}

\end{document}